\documentclass[twocolumn]{aastex63}
\usepackage{graphicx}
\usepackage{amsmath}
\usepackage{cases}
\usepackage{color}

\begin{document}

\title{Detection of PAH Absorption and Determination of the Mid-Infrared Diffuse Interstellar Extinction Curve from the Sightline Toward Cyg~OB2-12}

\author[0000-0001-7449-4638]{Brandon S. Hensley}
\email{bhensley@astro.princeton.edu}
\affiliation{Department of Astrophysical Sciences,  Princeton
  University, Princeton, NJ 08544, USA}
\affiliation{Spitzer Fellow}

\author[0000-0002-0846-936X]{B. T. Draine}
\affiliation{Department of Astrophysical Sciences,  Princeton
  University, Princeton, NJ 08544, USA}

\date{\today}

\begin{abstract}
    The sightline toward the luminous blue hypergiant Cyg~OB2-12 is widely used in studying interstellar dust on account of its large extinction ($A_V \simeq 10$\,mag) and the fact that this extinction appears to be dominated by dust typical of the diffuse interstellar medium. We present a new analysis of archival {\it ISO}-SWS and {\it Spitzer} IRS observations of Cyg~OB2-12 using a model of the emission from the star and its stellar wind to determine the total extinction $A_\lambda$ from 2.4--37\,$\mu$m. In addition to the prominent 9.7 and 18\,$\mu$m silicate features, we robustly detect absorption features associated with polycyclic aromatic hydrocarbons (PAHs), including the first identification of the 7.7\,$\mu$m feature in absorption. The 3.3\,$\mu$m aromatic feature is found to be much broader in absorption than is typically seen in emission. The 3.4 and 6.85\,$\mu$m aliphatic hydrocarbon features are observed with relative strengths consistent with observation of these features on sightlines toward the Galactic Center. We identify and characterize more than sixty spectral lines in this wavelength range, which may be useful in constraining models of the star and its stellar wind. Based on this analysis, we present an extinction curve $A_\lambda/A_{2.2\,\mu{\rm m}}$ that extrapolates smoothly to determinations of the mean Galactic extinction curve at shorter wavelengths and to dust opacities inferred from emission at longer wavelengths, providing a new constraint on models of interstellar dust in the mid-infrared.
\end{abstract}


\section{Introduction}

Vibrational modes in the molecular constituents of interstellar grains give rise to a number of infrared features seen in extinction and emission. Notable among these are the infrared emission features attributed to polycyclic aromatic hydrocarbons \citep[PAHs,][]{Leger+Puget_1984,Allamandola+Tielens+Barker_1985}, some of which have also been detected in extinction \citep[e.g.,][]{Schutte+etal_1998,Chiar+etal_2013}, and the prominent 9.7 and 18\,$\mu$m features arising from the Si-O stretching mode and O-Si-O bending mode in amorphous silicates, respectively \citep{Woolf+Ney_1969,vanBreemen+etal_2011}. Because of their identification with specific materials, these features enable detailed study of the chemical composition of grains as well as composition-specific properties, such as alignment efficiency.

Precise measurement of extinction by dust in the diffuse interstellar medium (ISM) at infrared wavelengths is an observational challenge. The mid-infrared (MIR) wavelengths central to this work are well into the Rayleigh-Jeans portion of stellar emission spectra, rendering most stars quite faint. Further, typical $\sim0.1\,\mu$m interstellar grains are far more efficient in extinguishing UV and optical radiation than infrared, resulting in a weak signal unless much dust is present. Most sightlines with high visual extinction $A_V$ pass through dense molecular material where the dust differs from that found in the diffuse ISM, due for instance to the presence of ice mantles.

The sightline toward the luminous blue hypergiant Cyg~OB2-12 (also known as Schulte~12) is frequently employed in studies of infrared extinction as it mitigates a number of these difficulties. One of the most intrinsically luminous stars in the Galaxy, Cyg~OB2-12 lies behind $\sim10$ magnitudes of visual extinction \citep{Humphreys_1978,TorresDodgen+etal_1991}, allowing infrared extinction to be measured with high signal to noise. Additionally, the bulk of the extinction appears to originate from dust in the diffuse ISM rather than dense gas, as evidenced by a typical extinction curve and a lack of strong ice features \citep{Whittet_2015}. Thus, this sightline has long been used to study the silicate features and the composition of interstellar silicates \citep{Rieke_1974,Adamson+Whittet+Duley_1990,Whittet+etal_1997,Schutte+etal_1998,Fogerty+etal_2016}.

Despite these advantages, the unusual nature of Cyg~OB2-12 also presents a number of modeling challenges. Photometric variability has been observed on short ($\lesssim$ 1 month) timescales in the X-ray \citep{Rauw_2011}, optical \citep{Naze+etal_2019}, and radio \citep{Scuderi+etal_1998,Morford+etal_2016}. A recently-discovered companion \citep{Caballero-Nieves+etal_2014,Maryeva+etal_2016}, though too faint to be detected in spectral features, raises the possibility of colliding stellar winds producing a significant fraction of the observed radio emission \citep{Oskinova+etal_2017}. Despite considerable effort modeling the panchromatic spectrum of Cyg~OB2-12 \citep[e.g.,][]{Clark+etal_2012}, fundamental uncertainties about the precise nature of this source and its emission remain \citep[see][for a recent discussion]{Naze+etal_2019}.

We focus in this work on modeling the MIR emission to derive the wavelength-dependent extinction from 2.4--37\,$\mu$m. Over this wavelength range, we employ physically-motivated parametric models of the stellar continuum emission and free-free emission from the stellar wind, as well as knowledge of the properties of interstellar dust, to infer the absolute interstellar extinction on this sightline. The resulting extinction curve, which includes prominent features associated with both silicate and carbonaceous grains, is presented as a benchmark for models of interstellar dust at MIR wavelengths.

This paper is organized as follows: in Section~\ref{sec:data}, we describe the data used in this work; in Section~\ref{sec:model}, we present the models of the star, stellar wind, and spectral lines employed; the resulting extinction curve is derived in Section~\ref{sec:results} and the identification and characterization of various extinction features is presented in Section~\ref{sec:features}; we discuss the implications of this work for dust models and potential directions for follow-up in Section~\ref{sec:discussion}; and finally, we summarize our principal conclusions in Section~\ref{sec:conclusions}.

\section{Data}
\label{sec:data}

\subsection{MIR Spectroscopy}
\subsection{ISO-SWS}
Cyg~OB2-12 was observed in three different epochs with the Short Wavelength Spectrometer (SWS) aboard the {\it Infrared Space Observatory} ({\it ISO}). Three Astronomical Observing Template 1 (AOT1) spectra are available from the NASA/IPAC Infrared Science Archive\footnote{\url{https://irsa.ipac.caltech.edu/data/SWS/}} having Target Dedicated Time numbers of 03602226, 13901048, and 33504130, corresponding to observations in December 1995, April 1996, and October 1996, respectively. These have been reduced as described in \citet{Sloan+etal_2003}. Because the detector type, noise properties, and relative agreement between these spectra all change longward of 4.08\,$\mu$m, for continuum measurements we analyze only the 2.35--4.08\,$\mu$m data, although longer wavelength {\it ISO} data are used in emission line characterization (see Section~\ref{subsec:lines}). For more details on the SWS, see \citet{deGraauw+etal_1996}. 

To improve the relative agreement between spectra, we multiply the 03602226 and 33504130 spectra by factors of 0.976 and 0.935, respectively. After this correction, the three spectra agree to within $\sim3$\% over the range of interest, with only small systematic trends with wavelength appearing at either end of the range.

The December 1995 spectrum (03602226), taken during {\it ISO} commissioning, has both the shortest duration at 18\,minutes and lowest spectral resolution \citep[$R \sim 200$;][]{Whittet+etal_1997}. While the other two spectra are of comparable duration ($\sim$1\,hour) and were taken at scan speed 3, affording higher spectral resolution, the April 1996 data (13901048) have more than twice the signal to noise over most of the wavelength range we consider. We therefore focus our analysis on this spectrum exclusively.

\subsubsection{Spitzer IRS}
Cyg~OB2-12 is one of 159 stars that compose the {\it Spitzer} Atlas of Stellar Spectra (SASS), a set of prototype stellar spectra taken by the {\it Spitzer} Infrared Spectrograph (IRS) that have been reduced in a homogeneous way \citep{Ardila+etal_2010}. The SASS Cyg~OB2-12 spectrum is a synthesis of two observations, one in October 2004 (AOR 9834496) and the other in August 2008 (AOR 27570176), that include data from both the Short-Low and High-Low modules \citep[for more details on the IRS, see][]{Houck+etal_2004}. Together, these provide full spectral coverage from 5.2 to 35\,$\mu$m. While we employ the IRS spectrum of Cyg~OB2-12 from SASS\footnote{\url{https://irsa.ipac.caltech.edu/data/SPITZER/SASS/fits/NAMEVICYG12\_matched.fits}}, \citet{Fogerty+etal_2016} find that their alternative reduction based on the same data is in overall good agreement with the SASS spectrum. We retain the full SASS spectrum in all plots, however we exclude data having $\lambda < 5.8\,\mu$m from our analysis since the spectrum in this region has relatively low signal to noise and differs qualitatively from the reduction of \citet{Fogerty+etal_2016}. 

The October 2004 IRS observation (AOR 9834496) also includes a high resolution spectrum ($R \sim 600$) from 10--37\,$\mu$m. We employ these data as obtained from the Cornell Atlas of {\it Spitzer}/Infrared Spectrograph Sources (CASSIS) using the optimal differential extraction \citep{Lebouteiller+etal_2015}. To ameliorate the discontinuity in flux densities between the short-high (SH) and long-high (LH) modules at 19.5\,$\mu$m, as well as to improve agreement with the SASS spectrum, we scale the SH data uniformly up by a factor of 1.06.

\subsection{UV, Optical, and Infrared Photometry}
\begin{deluxetable*}{lcccccr}
  \tablewidth{0pc}
      \tablecaption{Cyg~OB2-12 Continuum Fluxes
        \label{table:iropuv_obs}}
      \tablehead{\multicolumn{7}{c}{Assorted UV--NIR Photometry}}
          \startdata
     Band & $\lambda$
      & $m_X$ & $F_\nu\left(m_X=0\right)$ &
      $F_\nu^{\rm model}$ & $A_\lambda$ &
      \multicolumn{1}{c}{Reference} \\ & [$\mu$m] & [mag] & $[{\rm
          Jy}]$ & $[{\rm Jy}]$ & [mag] & \\
        $U$ & 0.366 & 17.15 & 1790 & 1000 & 16.52 & \citet{Wisniewski_etal_1967} \\
        $B$ & 0.438 & 14.70 & 4063 & 945 & 13.12 & \citet{Wisniewski_etal_1967} \\
        $V$ & 0.545 & 11.48 & 3636 & 816 & 9.86 & \citet{Wisniewski_etal_1967} \\
        $R$ & 0.641 & 8.26 & 3064 & 702 & 6.66 & \citet{Wisniewski_etal_1967} \\
        $I$ & 0.798 & 5.95 & 2416 & 547 & 4.34 & \citet{Wisniewski_etal_1967} \\
        $J$ & 1.22 & 4.38 & 1589 & 308 & 2.60 & \citet{Rieke+Lebofsky_1985} \\
        $H$ & 1.63 & 3.28 & 1021 & 200 & 1.51 & \citet{Rieke+Lebofsky_1985} \\
        $K$ & 2.19 & $2.715\pm0.02$ & 640 & 127 & 0.96 & \citet{Harris+etal_1978} \\
        $L$ & 3.45 & 2.217 & 285 & 63.2 & 0.58 & \citet{Harris+etal_1978} \\
        $L^\prime$ & 3.80 & 2.17 & 238 & 54.5 & 0.57 & \citet{TorresDodgen+etal_1991} \\ \hline \hline
        \multicolumn{7}{c}{APASS Photometry \citep{APASS_DR9}} \\ \hline
         Band & $\lambda$
      & $m_X$ & $F_\nu\left(m_X=0\right)$ &
      $F_\nu^{\rm model}$ & $A_\lambda$ & \\ & [$\mu$m] & [mag] & $[{\rm
          Jy}]$ & $[{\rm Jy}]$ & [mag] & \\
          $B$ & 0.438 & 14.929 & 4063 & 945 & 13.35 & \\
          $g$ & 0.4770 & 13.44 & 3631 & 900 & 11.93 \\
          $V$ & 0.545 & 11.58 & 3636 & 816 & 9.96 & \\
          $r$ & 0.6231 & 10.196 & 3631 & 722 & 8.44 \\ \hline \hline
             \multicolumn{7}{c}{{\it Gaia} Photometry \citep{Gaia_DR2} } \\ \hline
     Band & $\lambda_0$
      & \multicolumn{2}{c}{Flux} &
      \multicolumn{2}{c}{Model Flux} & \multicolumn{1}{c}{$A_{\lambda_0}$} \\ & [nm] & \multicolumn{2}{c}{[photoelectrons
        s$^{-1}$]} & \multicolumn{2}{c}{[photoelectrons
        s$^{-1}$]} & \multicolumn{1}{c}{[mag]} \\
        $G_{\rm BP}$ & 513.11 & \multicolumn{2}{c}{$3.11\times10^5$} & \multicolumn{2}{c}{$3.40\times10^9$} & \multicolumn{1}{c}{10.96} \\
        $G$ & 640.50 & \multicolumn{2}{c}{$5.25\times10^6$} & \multicolumn{2}{c}{$4.28\times10^9$} & \multicolumn{1}{c}{8.14} \\
        $G_{\rm RP}$ & 777.76 & \multicolumn{2}{c}{$8.48\times10^6$} & \multicolumn{2}{c}{$1.76\times10^9$} & \multicolumn{1}{c}{5.98} \\ \hline \hline
    \multicolumn{7}{c}{NIR--MIR Photometry from \citet{Leitherer+etal_1982}} \\ \hline
     Band & $\lambda$
      & $m_X$ & $F_\nu\left(m_X=0\right)$ &
      $F_\nu^{\rm model}$ & $A_\lambda$ &
      \\ & [$\mu$m] & [mag] & $[{\rm
          Jy}]$ & $[{\rm Jy}]$ & [mag] & \\
        $H$ & 1.67 & $3.33\pm0.01$ & 1076 & 193 & 1.46 &  \\
        $K$ & 2.30 & $2.72\pm0.01$ & 598 & 118 & 0.96 &  \\
        $L$ & 3.57 & $2.28\pm0.02$ & 277 & 60.0 & 0.62 &  \\
        $M$ & 4.97 & $2.06\pm0.02$ & 158 & 36.0 & 0.45 &  \\
        $N$ & 10.9 & $1.95\pm0.07$ & 33 & 9.65 & 0.61 &
    \enddata
    \tablenotetext{}{$UBVRIJHKLL'$ bandpass parameters taken from
      \citet{Bessell+etal_1998}.}
\end{deluxetable*}

Table~\ref{table:iropuv_obs} presents a heterogeneous collection of photometric observations of Cyg~OB2-12 from the literature. We adopt a representative set of $UBVRIJHKLL'$ photometry from \citet{Wisniewski_etal_1967}, \citet{Rieke+Lebofsky_1985}, \citet{Harris+etal_1978}, and \citet{TorresDodgen+etal_1991}. A more complete compilation of historical data can be found in \citet{Clark+etal_2012}. We note that \citet{Naze+etal_2019} demonstrate optical variability of Cyg~OB2-12 in the V band at the 0.1\,mag level on a one year timescale using data from ASAS-SN \citep{Kochanek_etal_2017} and a private observatory. Thus, multi-epoch comparisons are subject to uncertainties in excess of the formal photometric errors.

We supplement these data with observations from the AAVSO Photometric All Sky Survey (APASS) DR9 \citep{APASS_DR9}, {\it Gaia} DR2 \citep{Gaia_DR2}, and \citet{Leitherer+etal_1982}. All data, including the adopted zero levels, are listed in Table~\ref{table:iropuv_obs}.

\subsection{Radio Observations}
\begin{deluxetable}{lccr}
  \tablewidth{0pc}
      \tablecaption{Cyg~OB2-12 Radio Fluxes
        \label{table:radio_obs}}
      \tablehead{$\lambda$
      & Date & $F_\nu$ & \multicolumn{1}{c}{Reference} \\ $[{\rm cm}]$ & & $[{\rm mJy}]$ & }
          \startdata
      0.7 & April 1995 & $22.9\pm0.6$ & \citet{Contreras+etal_1996} \\
        0.7 & June 1999 & $9.0\pm1.5$ &
        \citet{Contreras+etal_2004} \\
        2 & April 1995 & $11.3\pm0.1$ & \citet{Contreras+etal_1996} \\
        2.1 & Sep. 1994 & $7.70\pm0.30$ & \citet{Scuderi+etal_1998} \\
        2.1 & Oct. 1994 & $12.0\pm0.20$ &
        \citet{Scuderi+etal_1998} \\
        3.5 & April 1995 & $7.18\pm0.04$ & \citet{Contreras+etal_1996} \\
        3.5 & Sep. 1994 & $4.74\pm0.14$ & \citet{Scuderi+etal_1998} \\
        3.5 & Oct. 1994 & $7.40\pm0.08$ &
        \citet{Scuderi+etal_1998} \\
        3.6 & May 1993 & $6.06\pm0.07$ &
        \citet{Waldron+etal_1998} \\
        3.6 & June 1999 & $5.9\pm0.1$ &
        \citet{Contreras+etal_2004} \\
        6 & April 1995 & $3.64\pm0.12$ & \citet{Contreras+etal_1996} \\
        6 & May 1993 & $3.94\pm0.07$ & \citet{Waldron+etal_1998}
        \\
        6 & June 1999 & $4.2\pm0.1$ & \citet{Contreras+etal_2004} \\
        6.2 & Sep. 1994 & $4.00\pm0.20$ & \citet{Scuderi+etal_1998} \\
        6.2 & Oct. 1994 & $5.03\pm0.10$ &
        \citet{Scuderi+etal_1998} \\
        21 & April 2014 & $1.013\pm0.055$ & \citet{Morford+etal_2016} \\
        21 & April 2014 & $0.598\pm0.061$ & \citet{Morford+etal_2016}
    \enddata
\end{deluxetable}

\begin{figure*}
\includegraphics[width=\textwidth]{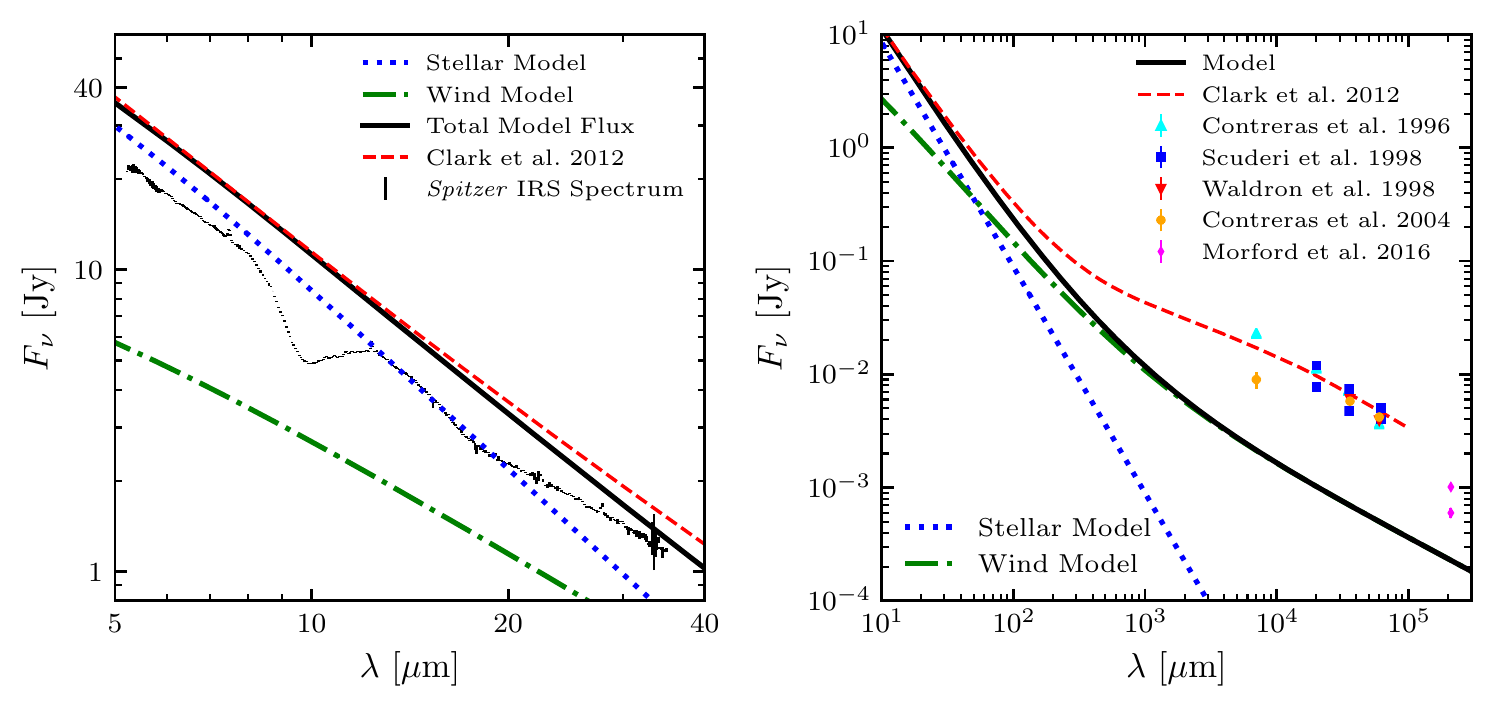}
\caption{\label{fig:CygOB2 model} In the left panel, we present our model of the flux from Cyg~OB2-12 (black solid), including contributions from the stellar disk (blue dotted) and stellar wind (green dot-dashed). The {\it Spitzer} IRS spectrum of Cyg OB-12 is plotted in black after removal of the emission lines (see Figure~\ref{fig:rlines2}). The \citet{Clark+etal_2012} model is plotted (red dashed) for comparison. In the right panel, we compare our model (black solid) and the \citet{Clark+etal_2012} model (red dashed) to observations of the radio continuum. We posit that free-free emission from the stellar wind of Cyg~OB2-12 is a subdominant component of the observed radio flux.
}
\end{figure*}

Cyg~OB2-12 has been studied extensively at radio wavelengths. Table~\ref{table:radio_obs} presents a selection of these observations, which are plotted in Figure~\ref{fig:CygOB2 model}. The presented data were taken both with the Very Large Array \citep{Contreras+etal_1996,Scuderi+etal_1998,Waldron+etal_1998,Contreras+etal_2004} and with e-MERLIN \citep{Morford+etal_2016}. Considerable variability is evident on both short ($\lesssim$ one month) and long ($\sim$years) timescales.

\section{Model}
\label{sec:model}

\subsection{Free-Free Emission from a Stellar Wind}
\label{sec:ff_wind}
The spectrum of a star with an ionized wind is discussed by \citet{Panagia+Felli_1975} and \citet{Wright+Barlow_1975}, and \citet{Clark+etal_2012} present a model for emission from Cyg~OB2-12. Our model builds on these results.

Let $R_0$ be the stellar radius, $D$ the distance, and $\theta_\star\equiv R_0/D$. We adopt the distance $D=1.75$\,kpc estimated by \citet{Clark+etal_2012}, which is consistent with the Gaia DR2 distances for the main Cygnus OB2 group \citep{Berlanas+etal_2019}. Cyg~OB2-12 has an effective temperature $T_{\rm eff}\approx 13700$\,K \citep{Clark+etal_2012}.

Assume that there is an ionized wind at $r>R_0$ with electron temperature $T_w$ and electron density $n_e(r)$. Let the total flux density from the star and wind be
\begin{equation}
F_\nu = F_\nu^\star + F_\nu^{\rm wind}
~~,
\end{equation}
where $F_\nu^\star$ is the flux density from the stellar disk (impact parameters $b<R_0$) and $F_\nu^{\rm wind}$ is the flux density from impact parameters $b>R_0$.  The ``stellar'' flux $F_\nu^\star$ includes the effects of radiative transfer through the wind projected in front of the stellar surface.

We assume the flux emerging from the photosphere to be a dilute blackbody,  with dilution factor $\epsilon_\star=0.75$ and temperature $T_\star=T_{\rm eff}\epsilon_\star^{-1/4}=14700$\,K. This dilution factor reproduces the MIR flux of an ATLAS9 model atmosphere having $T_{\rm eff}=14000$\,K and $g=10^2$\,cm\,s$^{-1}$ \citep{Castelli+Kurucz_2003}.

Following \citet{Clark+etal_2012}, we consider steady mass-loss $\dot{M}$ with electron density

\begin{align} \label{eq:density law}
n_e(r) &= n_{e0} \left(\frac{R_0}{r}\right)^2\left(1-0.913{\frac{R_0}{r}}\right)^{-3} \\
n_{e0} &\equiv \frac{\dot{M}/1.4m_{\rm H}}{4\pi R_0^2 v_\infty}
\end{align}
and adopt $v_\infty=400$\,km\,s$^{-1}$. At infrared and far-infrared wavelengths, the dominant opacity in the wind is free-free absorption, with attenuation coefficient
\begin{align}
\kappa_\nu^{\rm ff} 
&= A n_e^2
\\
A &=3.69\times10^{-21} T_4^{-1/2} \nu_9^{-3}
g_{\rm ff}(\nu)
\left[1-e^{-h\nu/kT_w}\right]\,{\rm cm}^5
~~,
\end{align}
where $T_4\equiv T_w/10^4$\,K, $\nu_9\equiv \nu/10^9$\,Hz, and $g_{\rm ff}(\nu)$ is the free-free Gaunt factor, approximated using Eq.\ (10.9) from \citet{Draine_2011}. We take $T_w \approx T_\star$. At impact parameter $b\equiv\beta R_0$ (with $\beta>1$) the optical depth in the wind is

\begin{align}
\tau_\nu(\beta\geq 1) &= \int_{-\infty}^{\infty} 
\kappa_\nu^{\rm ff}\left(r=\sqrt{\beta^2R_0^2+x^2}\right) dx \\
&= A n_{e0}^2 R_0 \int_{-\infty}^{\infty} \frac{\left(\beta^2 + u^2\right)du}{\left(\sqrt{\beta^2 + u^2} - 0.913\right)^6}
~~.
\end{align}
For an isothermal wind,

\begin{equation}
F_\nu^{\rm wind}(\nu) = B_\nu (T_w) \pi \theta_\star^2 
\int_1^\infty 2\beta d\beta \left[1 - e^{-\tau_\nu(\beta)}\right]
~~.
\end{equation}

For impact parameter $b < R_0$, the median optical depth in front of the stellar disk is
\begin{align}
\tau_\nu^m 
&=\int_0^\infty 
\kappa_\nu^{\rm ff}\left(r=\sqrt{R_0^2+x^2+xR_0\sqrt{2}}\right) dx
\\
&= 5.911\times10^4 A n_{e0}^2 R_0 
\end{align}
for the density profile (\ref{eq:density law}). We take

\begin{equation}
F_\nu^\star \approx \pi \theta_\star^2 B_\nu(T_\star)
\left[
\epsilon_\star e^{-\tau_\nu^m} + (1-e^{-\tau_\nu^m})
\right]
~~~.
\end{equation}

With this formulation, the continuum emission from the star and wind is set by only two free parameters: the angular size of the star $\theta_\star$ and the mass loss rate $\dot{M}$. We determine best-fit values of these parameters in Section~\ref{sec:results} on the basis of the resulting extinction curve.

\subsection{Line Emission}
\label{subsec:lines}

\begin{figure*}
\centering
\includegraphics[width=\textwidth]{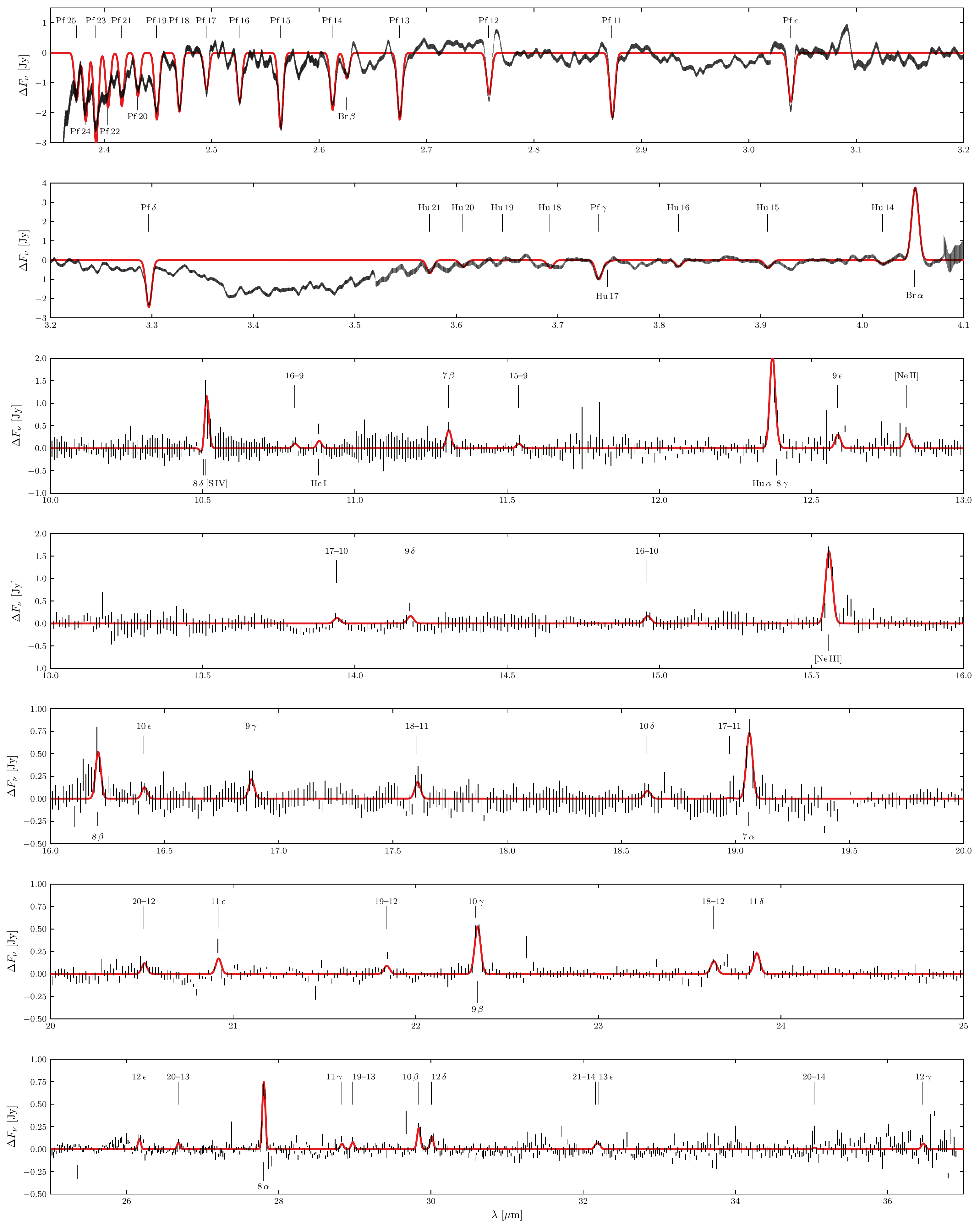}
\caption{Absorption and emission lines seen in the ISO-{\it SWS} spectrum (top two panels) and the high resolution {\it Spitzer} IRS spectrum (bottom five panels). A simple model fit employing Gaussian line profiles is presented in red, with the corresponding equivalent widths listed in Table~\ref{table:lines}.}
\label{fig:rlines2}
\end{figure*}

\begin{figure}
\centering
\includegraphics[width=\columnwidth]{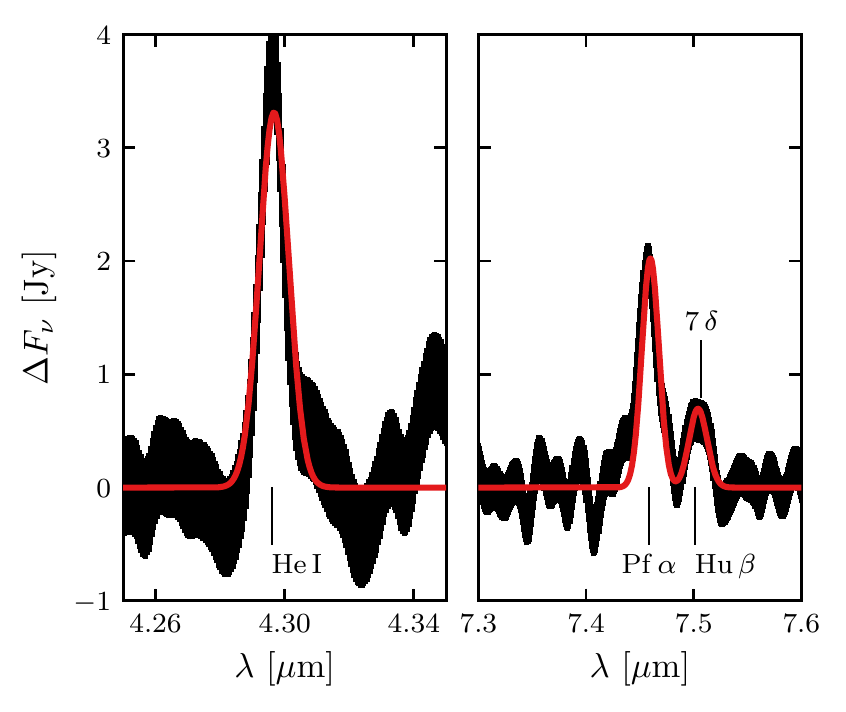}
\caption{{\it ISO}-SWS spectrum relative to the continuum in the vicinity of the He\,I 4.296\,$\mu$m and Pf\,$\alpha$ emission lines (left and right panels, respectively). The subdominant emission feature at 7.50--7.51\,$\mu$m is a combination of the Hu\,$\beta$ and $7\delta$ lines. Equivalent widths corresponding to the model fit (red) are given in Table~\ref{table:lines}.}
\label{fig:pfalpha}
\end{figure}

\begin{deluxetable*}{lcccc|lcccc}
  \tablewidth{0pc}
      \tablecaption{Cyg~OB2-12 Spectral Lines
        \label{table:lines}}
      \tablehead{\multicolumn{5}{c}{{\it ISO}-SWS} & \multicolumn{5}{c}{{\it Spitzer} IRS}}
          \startdata
     Name & $\lambda_0$
      & $n_u$ & $n_l$ &
      $W_\lambda$ &Name & $\lambda_0$
      & $n_u$ & $n_l$ &
      $W_\lambda$ \\
      & [$\mu$m] & & & [\AA] & & [$\mu$m] & & & [\AA] \\
Pf\,25 & 2.374 & 25 & 5 & $1.55$ & 8$\delta$ & 10.502 & 12 & 8 & $11^\dagger$ \\
Pf\,24 & 2.382 & 24 & 5 & $2.26$ & [S\,IV] & 10.511 &  & & $-46^\dagger$ \\
Pf\,23 & 2.392 & 23 & 5 & $3.03$ & 16\hbox{--}9 & 10.802 & 16 & 9 & $-3$ \\
Pf\,22 & 2.403 & 22 & 5 & $1.84$ & He\,I & 10.881 & 1s4p & 1s4s & $-6$ \\
Pf\,21 & 2.416 & 21 & 5 & $1.80$ & 7$\beta$ & 11.307 & 9 & 7 & $-15$ \\
Pf\,20 & 2.431 & 20 & 5 & $1.49$ & 15\hbox{--}9 & 11.537 & 15 & 9 & $-3$ \\
Pf\,19 & 2.449 & 19 & 5 & $2.31$ & Hu\,$\alpha$ & 12.370 & 7 & 6 & $-84^\dagger$ \\
Pf\,18 & 2.470 & 18 & 5 & $2.06$ & 8$\gamma$ & 12.385 & 11 & 8 & $-4^\dagger$ \\
Pf\,17 & 2.495 & 17 & 5 & $1.31$ & 9$\epsilon$ & 12.585 & 14 & 9 & $-12$ \\
Pf\,16 & 2.526 & 16 & 5 & $1.80$ & [Ne\,II] & 12.813 &  & & $-14$ \\
Pf\,15 & 2.564 & 15 & 5 & $2.81$ & 17\hbox{--}10 & 13.939 & 17 & 10 & $-6$ \\
Pf\,14 & 2.612 & 14 & 5 & $2.21$ & 9$\delta$ & 14.181 & 13 & 9 & $-9$ \\
Br\,$\beta$ & 2.625 & 6 & 4 & $1.01$ & 16\hbox{--}10 & 14.960 & 16 & 10 & $-11$ \\
Pf\,13 & 2.675 & 13 & 5 & $2.69$ & [Ne\,III] & 15.555 &  & & $-125$ \\
Pf\,12 & 2.758 & 12 & 5 & $1.76$ & 8$\beta$ & 16.206 & 10 & 8 & $-46$ \\
Pf\,11 & 2.872 & 11 & 5 & $2.97$ & 10$\epsilon$ & 16.409 & 15 & 10 & $-12$ \\
Pf\,$\epsilon$ & 3.039 & 10 & 5 & $2.52$ & 9$\gamma$ & 16.878 & 12 & 9 & $-22$ \\
Pf\,$\delta$ & 3.296 & 9 & 5 & $4.44$ & 18\hbox{--}11 & 17.605 & 18 & 11 & $-22$ \\
Hu\,21 & 3.573 & 21 & 6 & $1.47$ & 10$\delta$ & 18.612 & 14 & 10 & $-12$ \\
Hu\,20 & 3.606 & 20 & 6 & $0.80$ & 17\hbox{--}11 & 18.975 & 17 & 11 & $-1$ \\
Hu\,19 & 3.645 & 19 & 6 & $0.05$ & 7$\alpha$ & 19.059 & 8 & 7 & $-104$ \\
Hu\,18 & 3.692 & 18 & 6 & $0.95$ & 20\hbox{--}12 & 20.511 & 20 & 12 & $-19$ \\
Pf\,$\gamma$ & 3.740 & 8 & 5 & $2.40$ & 11$\epsilon$ & 20.917 & 16 & 11 & $-29$ \\
Hu\,17 & 3.749 & 17 & 6 & $0.19$ & 19\hbox{--}12 & 21.838 & 19 & 12 & $-17$ \\
Hu\,16 & 3.819 & 16 & 6 & $0.82$ & 10$\gamma$ & 22.328 & 13 & 10 & $-16^\dagger$ \\
Hu\,15 & 3.907 & 15 & 6 & $1.08$ & 9$\beta$ & 22.336 & 11 & 9 & $-90^\dagger$ \\
Hu\,14 & 4.020 & 14 & 6 & $0.56$ & 18\hbox{--}12 & 23.629 & 18 & 12 & $-31$ \\
Br\,$\alpha$ & 4.052 & 5 & 4 & $-10.62$ & 11$\delta$ & 23.864 & 15 & 11 & $-52$ \\
He\,I & 4.296 & 1s3p & 1s3s & $-14.15$ & 12$\epsilon$ & 26.164 & 17 & 12 & $-31$ \\
Pf\,$\alpha$ & 7.459 & 6 & 5 & $-31.45$ & 20\hbox{--}13 & 26.677 & 20 & 13 & $-21$ \\
Hu\,$\beta$ & 7.501 & 8 & 6 & $-8.03^\dagger$ & 8$\alpha$ & 27.798 & 9 & 8 & $-241$ \\
7$\delta$ & 7.507 & 11 & 7 & $-3.54^\dagger$ & 11$\gamma$ & 28.826 & 14 & 11 & $-21$ \\
& & & & & 19\hbox{--}13 & 28.967 & 19 & 13 & $-27$ \\
& & & & & 10$\beta$ & 29.834 & 12 & 10 & $-89$ \\
& & & & & 12$\delta$ & 30.005 & 16 & 12 & $-54$ \\
& & & & & 21\hbox{--}14 & 32.161 & 21 & 14 & $-20^\dagger$ \\
& & & & & 13$\epsilon$ & 32.204 & 18 & 13 & $-23^\dagger$ \\
& & & & & 20\hbox{--}14 & 35.034 & 20 & 14 & $-9$ \\
& & & & & 12$\gamma$ & 36.464 & 15 & 12 & $-41$ \\
    \enddata
    \tablecomments{Equivalent widths $W_\lambda$ correspond to the model in Figure~\ref{fig:rlines2}, with positive $W_\lambda$ denoting absorption lines and negative $W_\lambda$ denoting emission lines. Blended lines are indicated with a dagger ($\dagger$) and have uncertain relative equivalent widths.}
\end{deluxetable*}

In addition to the continuum emission from the star and stellar wind, a number of hydrogen recombination lines and other spectral lines contribute to the observed IR spectrum. Using {\it ISO} data, \citet{Whittet+etal_1997} identified Pf4--18 in absorption as well as Pfund\,$\alpha$ and Br\,$\alpha$ in emission. Humphreys\,$\alpha$ at 12.37\,$\mu$m has also been seen in emission toward Cyg~OB2-12 \citep{Bowey+Adamson+Whittet_1998,Fogerty+etal_2016}. Particularly in the high-resolution {\it Spitzer} IRS data, we detect a number of additional lines, and so we attempt to characterize them here.

Constructing a physical description of the line emission from the stellar wind would require a full non-LTE radiative transfer model due to the free-free, bound-free, and bound-bound opacity of the wind. Such a treatment is beyond the scope of our present study, and so we instead follow a simpler approach. First, we estimate the underlying continuum in each spectrum by performing a spline fit to the data between the lines. We subtract this continuum from the data to yield the $\Delta F_\nu$ presented in Figure~\ref{fig:rlines2}. For most lines, uncertainty in the continuum determination is the largest source of error in the computed line strength, but can be estimated at least roughly from the scatter around zero in Figure~\ref{fig:rlines2}.

After continuum subtraction, each line is modeled with a Gaussian profile having a FWHM set by the resolving power of the instrument: FWHM $\approx \lambda$/600 for the high resolution IRS spectrum and $\lambda$/625 for the AOT1 {\it ISO}-SWS spectrum at wavelengths $\lambda < 4.08\,\mu$m, consistent with the SWS resolution using AOT1 at scanning speed 3 \citep{deGraauw+etal_1996}. Finally, we estimate the strength of each line by performing a maximum likelihood fit to all lines simultaneously. The results of these fits are presented in Figure~\ref{fig:rlines2}, with the best fit equivalent widths $W_\lambda$ given in Table~\ref{table:lines}. Lines appear in absorption for $\lambda < 4.03\,\mu$m and generally in emission for $\lambda > 4.05\,\mu$m.

The {\it ISO}-SWS spectrum, particularly the Pfund series absorption lines, is best fit assuming a systematic redshift of 50\,km\,s$^{-1}$, which we apply uniformly to all line fits. A systematic redshift rather than blueshift is unexpected; detailed modeling of the line profiles provide a way to constrain the velocity of the star and the structure of its stellar wind, but we do not pursue this analysis here. We note that the heliocentric radial velocity of the Cyg~OB2 association is approximately $-11\,$km\,s$^{-1}$ \citep[][and references therein]{Klochkova+Chentsov_2004}.

As noted above, the Pf\,$\alpha$ 7.459\,$\mu$m line is prominent in emission on this sightline (see Figure~\ref{fig:CygOB2 model}), but is at a wavelength shorter than covered in the high resolution IRS spectrum and longer than where there is good mutual agreement between the three {\it ISO}-SWS spectra. Likewise, there is evidence for a strong emission line at 4.296\,$\mu$m which we attribute to He\,I. Since determining the line strengths is relatively insensitive to the calibration of the continuum, we fit these emission lines using the {\it ISO} data assuming line FWHMs of $\lambda/500$. 

\begin{figure*}
\includegraphics[width=\textwidth]{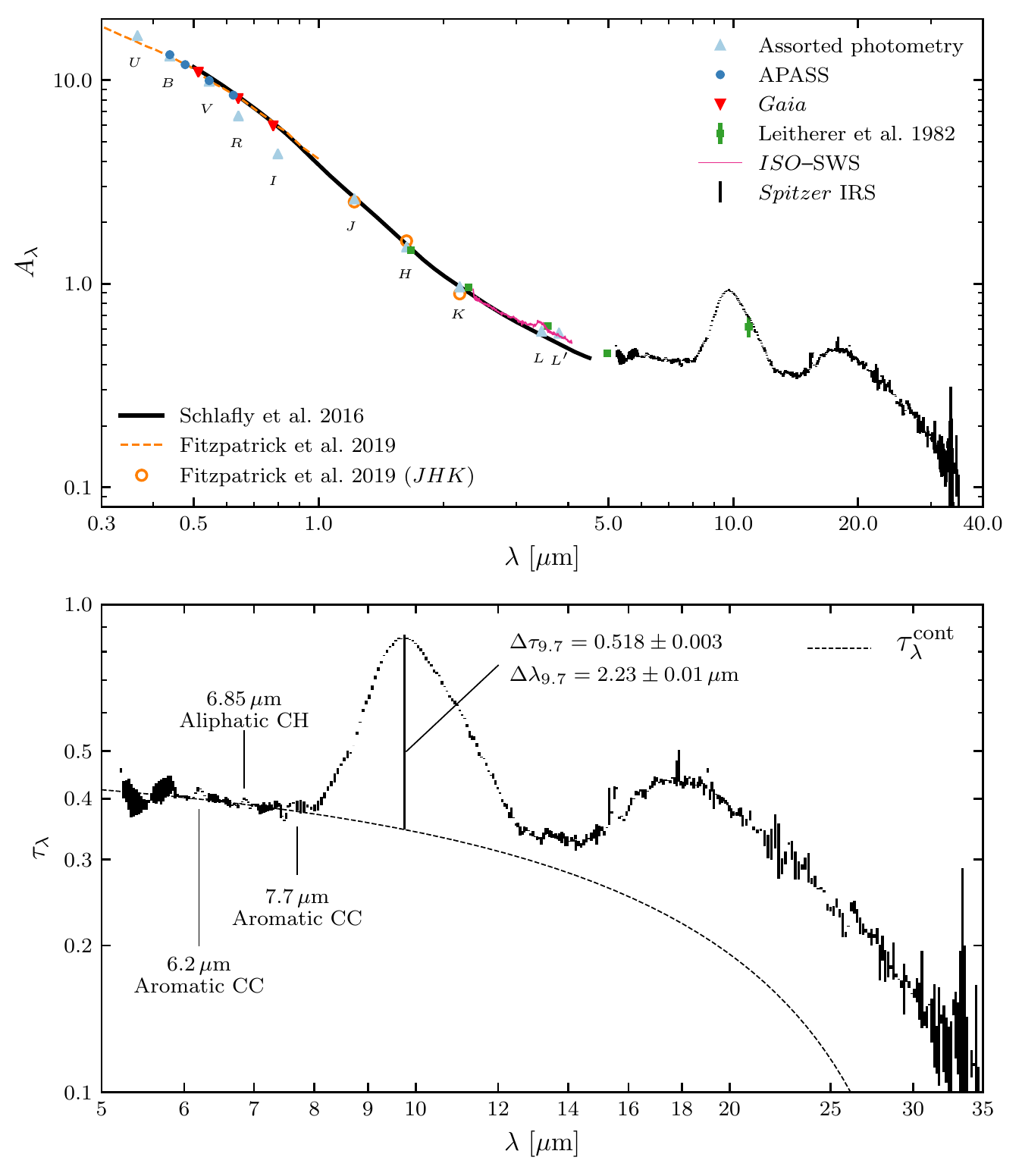}
\caption{\label{fig:tau} In the top panel, we plot the extinction curve computed from the modeled and observed fluxes (see Table~\ref{table:iropuv_obs}). We compare this curve to the Milky Way mean extinction curves derived by \citet{Schlafly+etal_2016} (black solid) and \citet{Fitzpatrick+etal_2019} (orange dashed). In the bottom panel, we plot the inferred optical depth $\tau_\lambda$ from the IRS spectrum. We find $\Delta\tau_{9.7} = 0.518$ using the adopted linear continuum (dashed line, Equation~\ref{eq:continuum}).
}
\end{figure*}

As shown in Figure~\ref{fig:pfalpha}, in addition to Pf\,$\alpha$ and He\,I, there is evidence for emission from Hu\,$\beta$ and/or $7\delta$ at 7.50 and 7.51\,$\mu$m, respectively. We note that a similar fit to the low resolution IRS data in the vicinity of the Pf\,$\alpha$ line yields line strengths compatible within the uncertainties of continuum determination and relative flux calibration, though with less ability to separate Pf\,$\alpha$ from the other emission lines. The difference is significant enough, however, that a residual remains when subtracting the SWS-inferred line fluxes from the IRS spectrum (visible in the bottom panel of Figure~\ref{fig:tau}), and so we excise this region of the spectrum when analyzing the IRS data in Section~\ref{subsec:carbon_features}.

We caution that the $W_\lambda$ presented in Table~\ref{table:lines} are intended to be estimates only. As we indicate in the table, some lines are indistinguishable at the spectral resolution of the data and so the contribution of each line to the total emission is difficult to discern. The transition between SH19 and SH20 in the IRS at 10.52\,$\mu$m occurs in the immediate vicinity of the $8\delta$ and $[{\rm S\,IV}]$ lines, rendering these lines strengths particularly uncertain. For most lines, the dominant uncertainty is in placement of the continuum, leading to typical variations in fit $W_\lambda$ of $\sim20$\% depending on modeling choices. These caveats notwithstanding, it is remarkable that nearly all of the $\alpha$, $\beta$, $\gamma$, and $\delta$ H recombination lines that fall within this wavelength range are clearly discernible in the spectra, with many transitions of even higher order visible as well.

The presence of both absorption and emission lines and an apparent non-monotonicity of line strength within a given spectral series attest to complicated line excitation physics. The wealth of information in these spectra can be used to constrain models like that developed in \citet{Clark+etal_2012} and elucidate the velocity profile, clumping, and other properties of the stellar wind.

In the following analysis of the low resolution IRS data, we subtract the contribution from the emission lines using the equivalent widths in Table~\ref{table:lines}. We assume a Gaussian profile for each line and adopt line FWHMs in each IRS module as recommended by PAHFIT \citep{Smith+etal_2007}. These FWHM values are 0.053\,$\mu$m for $\lambda < 7.55$\,$\mu$m, 0.10\,$\mu$m for $7.55 \leq \lambda/\mu$m $< 14.6$, 0.14\,$\mu$m for $14.6 \leq \lambda/\mu$m $< 20.7$, and 0.34\,$\mu$m for $\lambda \geq 20.7$\,$\mu$m. In this way we mitigate potential confusion between features induced by line emission versus dust extinction.

\section{The Mid-Infrared Extinction Toward Cyg~OB2-12}
\label{sec:results}

\subsection{Model Constraints}
\label{subsec:model_constraints}
After subtracting the best-fit models of the various lines, the continuum flux can be modeled using the formalism described in Section~\ref{sec:ff_wind}. To constrain the two free model parameters $\theta_\star$ and $\dot{M}$, we consider what is known about the shape of the extinction curve.

Despite the large amount of reddening, the extinction along the sightline toward Cyg~OB2-12 appears typical of the diffuse ISM in both the shape of the UV/optical extinction curve and the lack of ice features \citep{Whittet_2015}. Thus, we posit that the extinction law toward Cyg~OB2-12 should also agree with recent determinations of the mean Galactic extinction curve at infrared wavelengths. 

Employing a sample of 37,000 stars and photometry from PAN-STARRS1, the Two Micron All-Sky Survey (2MASS), and the {\it Wide-field Infrared Survey Explorer} (WISE), \citet{Schlafly+etal_2016} made a determination of the mean Galactic interstellar extinction curve extending from 5000\,\AA\ to 4.5\,$\mu$m. This curve is shown in the top panel of Figure~\ref{fig:tau} for its default parameters of $A_H/A_K = 1.55$ \citep{Indebetouw+etal_2005} and $x = 0$ (equivalent to $R_V \simeq 3.3$). 
Recently, \citet{Fitzpatrick+etal_2019} derived the mean Milky Way interstellar extinction law from the ultraviolet to the near-infrared using spectrophotometry from the {\it Hubble Space Telescope}, archival data from the {\it International Ultraviolet Explorer}, and photometry in the $JHK$ bands from 2MASS. We use this extinction law to guide our model fits as well and present it alongside that of \citet{Schlafly+etal_2016} in Figure~\ref{fig:tau}.

At far-infrared (FIR) wavelengths, the properties of interstellar dust are well-constrained by observations of dust {\it emission}. In particular, {\it Planck} observations of dust emission are well-fit by a power law dust opacity $\kappa_\nu$, where $\kappa_\nu \propto \lambda^{-1.53}$ between 350\,$\mu$m and $\sim$3\,mm \citep{Planck_Int_XXII, Planck_2015_X}. Since scattering is negligible at these wavelengths, the extinction cross section and thus the optical depth should also scale as $\lambda^{-1.53}$ for $\lambda \gtrsim 350\,\mu$m. We therefore require our model to yield an extinction curve having approximately this behavior longward of the 18\,$\mu$m silicate feature.

We can in principle improve further on the connection of MIR extinction to FIR emission. The FIR dust emission is well-fit with a dust temperature of 20\,K \citep{Planck_Int_XXII}, and the dust emission in the {\it Planck} 857\,GHz (350\,$\mu$m) band per $N_{\rm H}$ at high Galactic latitudes has been determined to be $4.3\times10^{-21}$\,MJy\,sr$^{-1}$\,cm$^2$ \citep{Planck_Int_XVII}. This implies $\tau_{857}/N_{\rm H} = 3.2\times10^{-26}$\,cm$^2$. On high-latitude sightlines, $N_{\rm H}/E(B-V) = 8.8\times10^{21}$\,cm$^{-2}$\,mag$^{-1}$ \citep{Lenz+Hensley+Dore_2017}, and so $\tau_{857}/A_V = 9.0\times10^{-5}$\,mag$^{-1}$ assuming $R_V = 3.1$. Thus, if the ratio of optical to infrared extinction is constant across the sky, we would expect the sightline toward Cyg~OB2-12 to have $\tau_{857} \simeq 9.0\times10^{-4}$ since $A_V \simeq 10$.

Alternatively, $N_{\rm H}$ has been estimated toward Cyg~OB2-12 to be $2\times10^{22}$\,cm$^{-2}$ by fitting X-ray absorption data \citep{Oskinova+etal_2017}. We note that determinations of $N_{\rm H}/E(B-V)$ near the Galactic plane \citep[e.g.,][]{Bohlin+Savage+Drake_1978} are $\sim50$\% lower than  those at high Galactic latitudes \citep[e.g.,][]{Liszt_2014a,Lenz+Hensley+Dore_2017}, suggesting a possible systematic difference in dust to gas ratio. If we assume that the interstellar dust toward Cyg~OB2-12 has the same properties as that observed at high Galactic latitudes but is simply 50\% more abundant per H atom, then we would estimate that $\tau_{857} \simeq 3.2\times10^{-26}$\,cm$^2\times2\times10^{22}$\,cm$^{-2}\times1.5 = 9.5\times10^{-4}$.

Given the concordance between these estimates, we require that our model yield an extinction curve which extrapolates to $\tau \simeq 9\times10^{-4}$ at 350\,$\mu$m. Using the dust emission per H atom measured at other frequencies by \citet{Planck_Int_XVII}, we likewise estimate $\tau \simeq 4\times10^{-4}$ at 550\,$\mu$m and $\tau \simeq 2\times10^{-4}$ at 850\,$\mu$m.

\subsection{Model Fits}
Using the formalism outlined in Section~\ref{sec:ff_wind}, we can compute a model flux at every wavelength of interest given values for $\theta_\star$ and $\dot{M}$. We obtain the extinction curve by comparing to the observed fluxes listed in Table~\ref{table:iropuv_obs}. For all data except from those from {\it Gaia}, we do not take into account the instrumental bandpasses, i.e., we assume the observed fluxes are the monochromatic fluxes at the central wavelength. However, the {\it Gaia} bandpasses are quite broad and so we consider the bandpasses explicitly. Specifically, we assume that the extinction law across each bandpass is given by the \citet{Fitzpatrick+etal_2019} curve and solve for total extinction at the nominal wavelength $\lambda_0$. We work in units of photoelectrons per second following the {\it Gaia} Data Release 2 Documentation \citep[v1.2;][]{Gaia_DR2_Doc}.

We find that $\theta_\star = 2.65\times10^{-9}$\,rad and $\dot{M} = 2.4\times10^{-6}\,M_\odot$\,yr$^{-1}$ produce an extinction curve most consistent with the considerations discussed in Section~\ref{subsec:model_constraints}. The model fluxes for the star and wind are illustrated in Figure~\ref{fig:CygOB2 model}. The resulting extinction curve is presented in detail in Figure~\ref{fig:tau}, and the extinction $A_\lambda$ in each photometric band is listed in Table~\ref{table:iropuv_obs}. 

Good agreement is obtained with both the \citet{Schlafly+etal_2016} and \citet{Fitzpatrick+etal_2019} mean extinction laws, with only the historical $R$ and $I$ band measurements being significantly discrepant. Given that this disagreement is not found with the {\it Gaia} observations of Cyg~OB2-12 over the same wavelength range, this might be due to the $R$ and $I$ bandpasses being significantly different than assumed or, particularly in light of the long time baseline, stellar variability. The minor difference between our derived $A_V = 9.86$ and other determinations of $A_V \simeq 10.2$ \citep[e.g.][]{Humphreys_1978,TorresDodgen+etal_1991,Clark+etal_2012} is not unexpected given our simple model of the stellar emission which, while consistent with more detailed modeling in the infrared, does not capture the complexities at optical and UV wavelengths \citep{Castelli+Kurucz_2003,Clark+etal_2012}.

With these parameters and a distance of 1.75\,kpc, Cyg~OB2-12 has a luminosity $L_\star=1.4\times10^6\,L_\odot$. Our adopted $\dot{M} = 2.4\times10^{-6}\,M_\odot$\,yr$^{-1}$ is close to the value $\dot{M}=3\times10^{-6}\,M_\odot\,$yr$^{-1}$ estimated by \citet{Clark+etal_2012}. In our model, the wind and the stellar disk contribute equally at 50\,$\mu$m.

While our model and that of \citet{Clark+etal_2012} are very similar over the wavelengths covered by the {\it Spitzer} IRS observations, the small differences are significant for our purposes. In particular, the \citet{Clark+etal_2012} model implies a significantly larger 30\,$\mu$m extinction, which is difficult to reconcile with the FIR opacities inferred from dust emission. Likewise, the implied 25--35\,$\mu$m extinction would then fall off too slowly compared to the $\sim\lambda^{-1.53}$ behavior expected in the FIR.

Figure~\ref{fig:CygOB2 model} presents a number of radio observations of Cyg OB2-2. \citet{Clark+etal_2012} assume clumping in the outer wind in order to reproduce the observed radio emission. However, recent detection of variability at 20\,cm over only 14 days \citep{Morford+etal_2016} suggests that some other source may be responsible for much of the flux at $\lambda > 6$\,cm, with the $\sim$400\,km\,s$^{-1}$ wind from Cyg~OB2-12 accounting for only a fraction of the observed radio emission.

There is an X-ray source coincident with Cyg~OB2-12 \citep{Waldron+etal_1998, Oskinova+etal_2017}. \citet{Oskinova+etal_2017} suggest that the X-ray emission may arise from colliding stellar winds, if the close companion recently discovered by \citet{Caballero-Nieves+etal_2014} is an O star with a fast wind. This colliding wind scenario could account for much of the observed radio emission, but should not affect the 5--35\,$\mu$m spectrum of interest here (except perhaps for emission lines from species such as [SIV]). Thus, we are unconcerned that our model flux is well below the observed radio emission. High angular resolution observations are needed to clarify the origin of the mm-wave continuum.

\subsection{Normalized Extinction Curve}
\label{subsec:extcrv}

\begin{figure*}
    \centering
        \includegraphics[width=\textwidth]{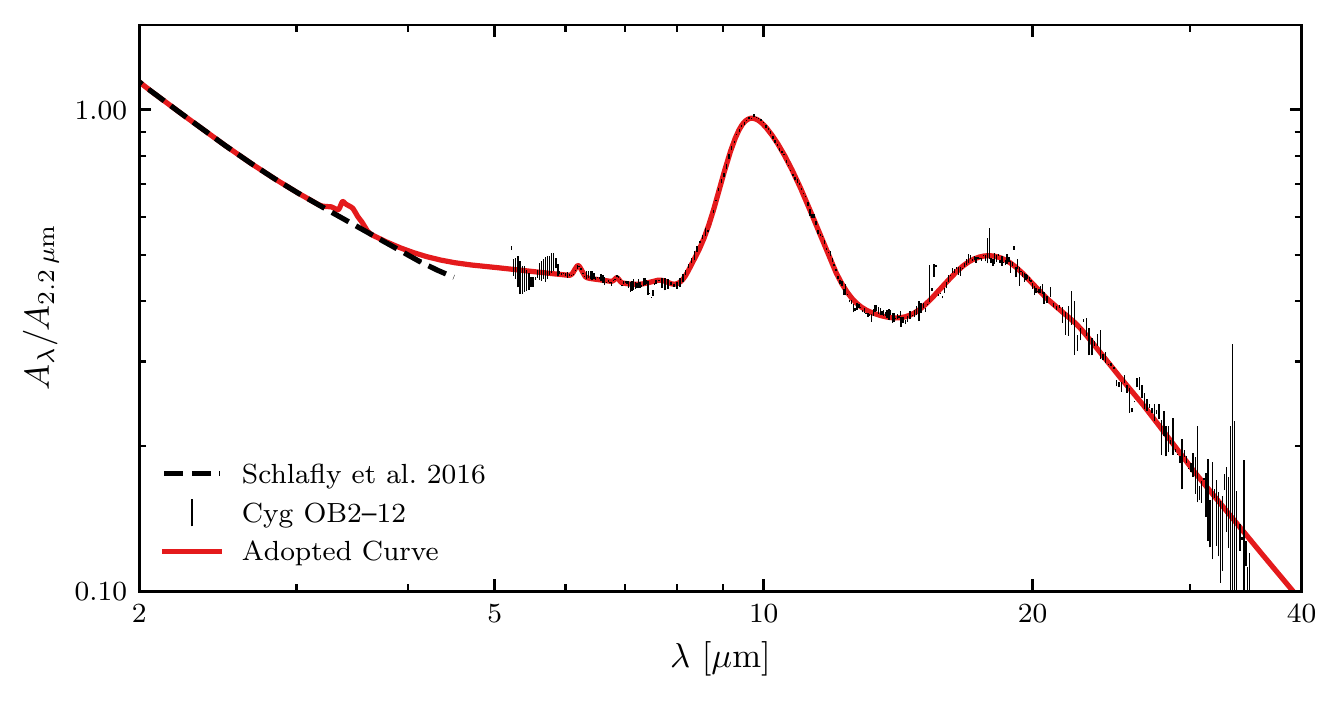}
    \caption{We construct the representative extinction curve presented in this work (red solid) by joining the mean Milky Way extinction curve in the optical and near-IR from \citet{Schlafly+etal_2016} (black dashed) with our determination of the total extinction toward Cyg~OB2-12 from the {\it Spitzer} IRS spectroscopy (black error bars). Extrapolated to FIR wavelengths, the adopted curve is also consistent with dust opacities inferred from {\it Planck} observations of dust emission (see Section~\ref{subsec:extcrv}). The 3.4\,$\mu$m feature is added to the \citet{Schlafly+etal_2016} curve following our determination from the {\it ISO}-SWS data (see Section~\ref{subsec:carbon_features}).} \label{fig:new_extcrv} 
\end{figure*}

\begin{figure}
    \centering
        \includegraphics[width=\columnwidth]{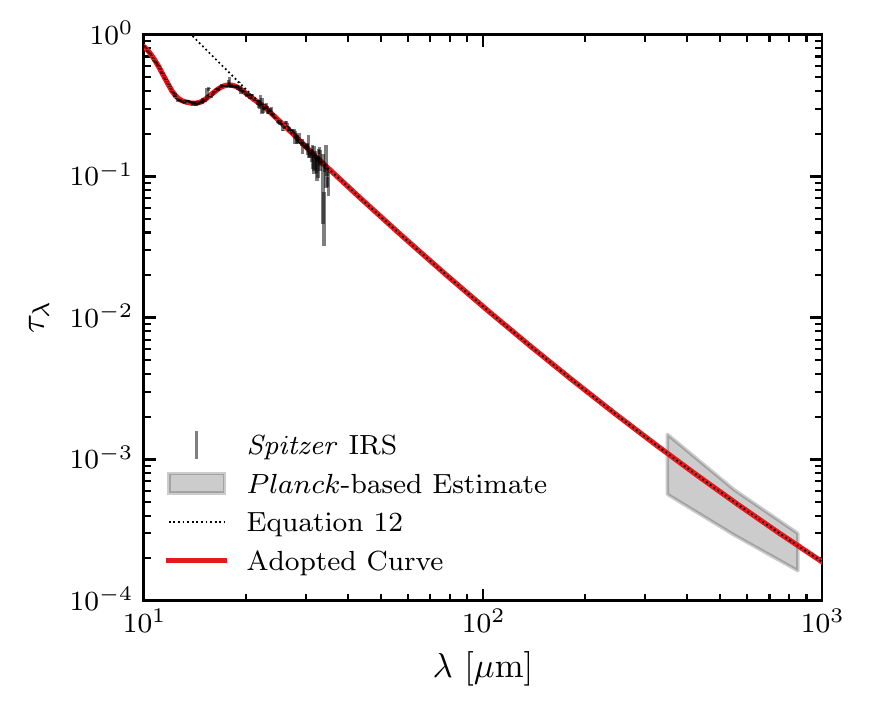}
    \caption{Comparison of the adopted extinction law to the {\it Spitzer} IRS spectrum of Cyg~OB2-12 (black error bars) and the $\tau_\lambda$ estimates made from {\it Planck} observations of dust emission (shaded region, see Section~\ref{subsec:model_constraints}). The extent of the shaded region represents varying the dust temperature from 16 to 24\,K. The parameterization given in Equation~\ref{eq:fir_tau} is shown by the dotted line.} \label{fig:planck} 
\end{figure}

With its high signal to noise and spectral resolution, the {\it Spitzer} IRS spectrum of Cyg~OB2-12 provides perhaps the most detailed characterization of MIR extinction from the diffuse ISM of any current data. We therefore propose using these data to extend determinations of the mean Galactic extinction curve into the MIR and even FIR.

In Figure~\ref{fig:new_extcrv}, we plot our Cyg~OB2-12 extinction curve normalized to unity at 2.2\,$\mu$m, roughly $K$ band, assuming $A_{2.2\,\mu{\rm m}} = 0.96$ (see Table~\ref{table:iropuv_obs}). We illustrate in red our synthesized curve, which matches onto the \citet{Schlafly+etal_2016} curve at short wavelengths and interpolates smoothly through the {\it Spitzer} IRS data in the MIR. Based on broadband photometry, the \citet{Schlafly+etal_2016} extinction law does not include spectral features, and so we employ the {\it ISO}-SWS spectrum to characterize this spectroscopic feature near 3.4\,$\mu$m (see Section~\ref{subsec:carbon_features} below).

The extinction at $\lambda \gtrsim 20\,\mu$m is well-fit by

\begin{equation}
\label{eq:fir_tau}
    A_\lambda/A_{2.2\,\mu{\rm m}} = 2.76\times10^{-4} \left(\frac{850\,\mu{\rm m}}{\lambda}\right)^{1.52 + \Delta\beta\left(\lambda\right)}
    ~~~,
\end{equation}
where

\begin{equation}
    \Delta\beta\left(\lambda\right) = 0.1\ln\left[\frac{2\,{\rm mm}}{{\rm min}\left(\lambda,2\,{\rm mm}\right)}\right]
\end{equation}
and $\ln$ is the natural logarithm. For Cyg~OB2-12, this yields $\tau = 0.41$ at 20\,$\mu$m, $1.1\times10^{-3}$ at 350\,$\mu$m, and $2.4\times10^{-4}$ at 850\,$\mu$m, in agreement with Figure~\ref{fig:tau} and the FIR estimates based on {\it Planck} data discussed in Section~\ref{subsec:model_constraints}. Further, the polarized dust intensity measured by {\it Planck} is well-described with an opacity scaling as $\lambda^{-1.53\pm0.02}$ from 850\,$\mu$m to 7.5\,mm \citep{Planck_2018_XI}, suggesting that Equation~\ref{eq:fir_tau} is an appropriate estimate well into the microwave. The extrapolation to the {\it Planck} frequencies is illustrated in Figure~\ref{fig:planck}, where the shaded band shows the effect of varying the assumed dust temperature between 16 and 24\,K.

To the extent that the sightline toward Cyg~OB2-12 typifies extinction from the diffuse ISM, our synthesized extinction curve extends the determinations of the mean Galactic extinction from $\sim4\,\mu$m through the FIR. We will make this extinction curve available in tabular form following publication.

\section{Extinction Features Toward Cyg~OB2-12}
\label{sec:features}
The high signal to noise and spectral resolution of the {\it ISO}-SWS and {\it Spitzer} IRS data enable identification of a number of spectroscopic extinction features that have been identified with specific materials. In this section, we identify and characterize a number of these features.

\begin{deluxetable}{cccccc}
  \tablewidth{0pc}
      \tablecaption{Cyg~OB2-12 Extinction Features 
        \label{table:features}}
      \tablehead{\multicolumn{5}{c}{Silicate Features}}
          \startdata
     $\lambda_0$ & $\Delta\tau$ &
      $\Delta\lambda$ & $\Delta\lambda^{-1}$ & $\int d\lambda^{-1} \Delta\tau$\\
      $[\mu$m] &  & [$\mu$m] & [cm$^{-1}$]
          &  [cm$^{-1}$]\\ 
          9.7 & $0.518\pm0.003$ & $2.23\pm0.01$ & $240\pm1$ & $119\pm2$ \\ 
          18 & 0.22 & 5.7 & 172 & 68 \\ \hline \hline
    \multicolumn{5}{c}{Carbonaceous Features} \\ \hline
     $\lambda_0$ & $\Delta\tau$ &
      $\Delta\lambda$ & $\Delta\lambda^{-1}$ & $\int d\lambda^{-1} \Delta\tau$ \\ 
      $[\mu$m] &  & [$\mu$m] & [cm$^{-1}$]
          &  [cm$^{-1}$]\\
          3.3 & $0.014\pm0.005$ & 0.10 & 96 & 1.41 \\
          3.4 & $0.044\pm0.005$ & 0.18 & 156 & 6.61  \\
          6.2 & $0.022\pm0.001$ & $0.127\pm0.003$ & $32.9\pm0.8$ & $0.78\pm0.03$ \\
          6.85 & $0.009\pm0.001$ & $0.086\pm0.009$ & $18\pm2$ & $0.21\pm0.03$ \\
          7.7 & $0.017\pm0.002$ & $0.54\pm0.08$ & $91\pm13$ & $2.5\pm0.4$ \\
    \enddata
    \tablecomments{Reported uncertainties on the 6.2, 6.85, 7.7, and 9.7\,$\mu$m feature parameters are statistical only. The properties of the 3.3 and 3.4\,$\mu$m features are derived from the best fit Gaussian decomposition presented in Figure~\ref{fig:iso34} and Table~\ref{table:iso_34} with quoted uncertainties estimated from alternate fits of the underlying continuum. Parameters of the 18\,$\mu$m feature are quoted based on the fiducial continuum model (Equation~\ref{eq:continuum}), which is relatively unconstrained at these wavelengths.}
\end{deluxetable}

\subsection{Continuum Extinction}
Before analyzing the profiles of the MIR dust extinction features, it is first necessary to determine the underlying continuum. Our estimate of the continuum is based on the 6--8\,$\mu$m IRS spectrum between the dust extinction features. At shorter wavelengths, the IRS data become noisy, and at longer wavelengths the extinction is dominated by the silicate features. In Section~\ref{subsec:carbon_features}, we demonstrate that the simple linear function

\begin{equation}
\label{eq:continuum}
    \tau_\lambda^{\rm cont} = -0.0155\left(\frac{\lambda}{\mu{\rm m}}\right) + 0.494
\end{equation}
describes the continuum extinction over the 6--8\,$\mu$m range and extrapolates well to the extinction curve at shorter wavelengths (see Figures~\ref{fig:tau} and \ref{fig:c_features}). The extinction in the dust features $\Delta\tau_\lambda$ is then determined by

\begin{equation}
    \Delta\tau_\lambda \equiv \tau_\lambda - \tau_\lambda^{\rm cont}
    ~~~.
\end{equation}

\subsection{Silicate Features}
As seen in Figure~\ref{fig:tau}, the most prominent MIR dust extinction features toward Cyg~OB2-12 are the 9.7 and 18\,$\mu$m silicate features. To determine the feature profiles, we model the underlying continuum with Equation~\ref{eq:continuum}. While it is probably reasonable to extrapolate Equation~\ref{eq:continuum} to 9.7\,$\mu$m, the assumption of a linear continuum becomes increasingly unreliable at longer wavelengths, with the adopted function eventually going to zero at 31.9\,$\mu$m.

Including the uncertainty in the underlying continuum from the fits in Section~\ref{subsec:carbon_features}, we find that $\Delta\tau_{9.7} = 0.518\pm0.003$. The feature has a FWHM of $2.23\pm0.01$\,$\mu$m and an integrated area of $119\pm2$\,cm$^{-1}$. We list these values in Table~\ref{table:features}. Assuming $A_V \simeq 10$, this implies $A_V/\Delta\tau_{9.7} = 19.3$, within the range observed on other sightlines \citep{Draine_2003}.

Extrapolating Equation~\ref{eq:continuum} to 18\,$\mu$m, we find that $\tau_{18}^{\rm cont}=0.22$ and $\Delta\tau_{18} = 0.22\pm0.01$ with $\Delta\tau_{18}/\Delta\tau_{9.7} \simeq 0.42$. Based on the short wavelength side of the feature only, we estimate a FWHM of 5.7\,$\mu$m, extending roughly from 15.6 to 21.2\,$\mu$m and peaking at 18.4\,$\mu$m. However, these quantities depend sensitively on the underlying continuum, which is relatively unconstrained particularly on the long wavelength side.

The detailed shapes of the silicate features provide constraints on the precise composition of interstellar silicate materials, such as the O:Si:Mg:Fe ratios. \citet{Fogerty+etal_2016} performed a detailed comparison of these data to laboratory materials, finding evidence for a silicate stoichiometry intermediate between olivine and pyroxene. While we do not perform additional analysis on the nature of the silicate material itself, we note that evident subfeatures in the profiles in Figure~\ref{fig:tau} that have persisted even after subtracting line emission from the stellar wind may provide additional clues to the detailed composition of interstellar silicates and should be further pursued. Of particular note is an apparent feature at $\sim 13.8\,\mu$m in the IRS data (see Figures~\ref{fig:rlines2} and \ref{fig:tau}), though it is unclear whether this is astrophysical rather than instrumental in origin.

\subsection{Carbonaceous Features}
\label{subsec:carbon_features}

A close inspection of the Cyg~OB2-12 extinction curve reveals absorption features in addition to the prominent silicate features, as indicated in Figure~\ref{fig:tau}. The {\it ISO}-SWS spectrum allows detailed characterization of the prominent 3.4\,$\mu$m feature associated with aliphatic hydrocarbons. The features in the IRS spectrum at 6.2 and 7.7\,$\mu$m are recognizable as PAH features, though seen in absorption rather than emission. In addition, we detect the 6.85\,$\mu$m feature arising from aliphatic hydrocarbons. We now explore each of these features in greater detail.

\subsubsection{The 3.4\,$\mu$m Complex}

\begin{figure*}
    \centering
        \includegraphics[width=\textwidth]{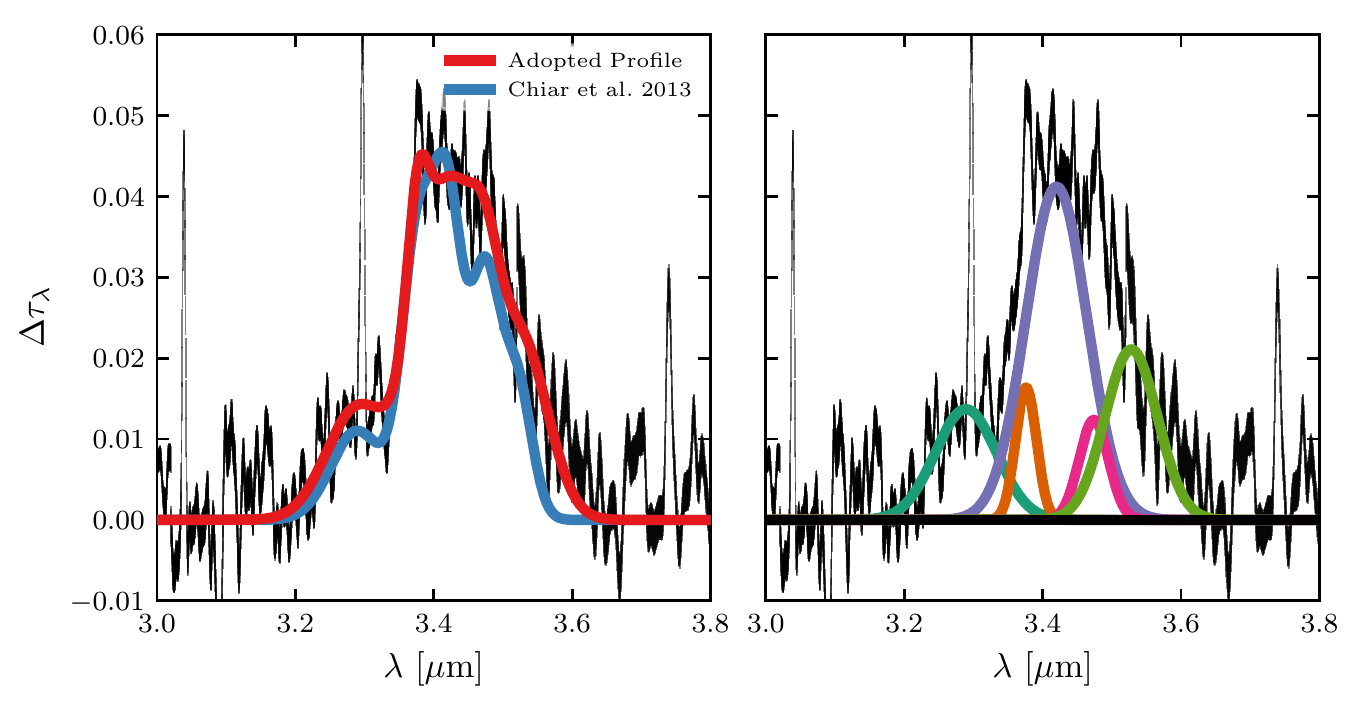}
    \caption{Determination of the 3.4\,$\mu$m feature profile from the {\it ISO}-SWS spectrum (black). An absorption feature at 3.3\,$\mu$m attributed to aromatic carbon is also present. In the left panel, we present the Gaussian decomposition of the feature profile in red, while the right panel shows each Gaussian component. The corresponding fit parameters listed in Table~\ref{table:iso_34}. The 3.4\,$\mu$m feature profile derived on sightlines toward the Quintuplet Cluster \citep{Chiar+etal_2013} is shown in the right panel in blue, demonstrating excellent agreement. The prominent absorption lines at 3.039, 3.296, and 3.740\,$\mu$m are Pf\,$\epsilon$, $\delta$, and $\gamma$, respectively.} \label{fig:iso34} 
\end{figure*}

\begin{deluxetable}{cccccc}
  \tablewidth{0pc}
      \tablecaption{Gaussian Decomposition of the 3.4\,$\mu$m Feature   
        \label{table:iso_34}}
      \tablehead{$\lambda_0$ & $\Delta\tau_{\lambda_0}$ &
      $\Delta\lambda$ & $\Delta\lambda^{-1}$ & $\int d\lambda^{-1} \Delta\tau$ \\ $[\mu$m] &  & [$\mu$m] & [cm$^{-1}$]
          &  [cm$^{-1}$]}
          \startdata
3.289 & 0.014 & 0.10 & 96 & 1.41 \\
3.376 & 0.016 & 0.04 & 31 & 0.55 \\
3.420 & 0.041 & 0.10 & 89 & 3.91 \\
3.474 & 0.012 & 0.05 & 41 & 0.55 \\
3.528 & 0.021 & 0.09 & 71 & 1.61 \\
    \enddata
\end{deluxetable}

By far the most prominent extinction feature visible in the {\it ISO}-SWS spectrum is the extinction feature at 3.4\,$\mu$m (see Figure~\ref{fig:tau}) arising from the C--H stretching mode in aliphatic hydrocarbons. Prior determinations of the strength of this feature toward Cyg~OB2-12 have been made with UKIRT \citep{Adamson+Whittet+Duley_1990} and {\it ISO} \citep{Whittet+etal_1997}, which found $\Delta\tau_{3.4} = 0.03\pm0.01$ and $0.04\pm0.01$, respectively.

We present our determination of the 3.4\,$\mu$m feature profile in Figure~\ref{fig:iso34}. The depth of the feature at 3.4\,$\mu$m depends on the details of the assumed continuum. We estimate $\Delta\tau_{3.4} = 0.044\pm0.005$, in good agreement with the $\Delta\tau_{3.4}$ derived by \citet{Whittet+etal_1997} but somewhat higher than that of \citet{Adamson+Whittet+Duley_1990}, whose determination has $\Delta\tau = 0$ near 3.3\,$\mu$m. We estimate a feature FWHM of 0.18\,$\mu$m.

Using {\it ISO}-SWS measurements toward the Quintuplet Cluster, \citet{Chiar+etal_2013} derived a Gaussian decomposition of the extinction near 3.4\,$\mu$m using five distinct components. We compare that profile to the observed extinction toward Cyg~OB2-12 in Figure~\ref{fig:iso34}, where we have scaled it to $\Delta\tau_{3.4} = 0.044$. The overall agreement is very good. In particular, both the \citet{Chiar+etal_2013} profile and the Cyg~OB2-12 spectrum suggest a feature at 3.3\,$\mu$m expected from aromatic hydrocarbons with strength $\Delta\tau_{3.3} \simeq 0.01$.

The \citet{Chiar+etal_2013} profile departs from the Cyg~OB2-12 spectrum in two principal ways. First, it underestimates the absorption in the vicinity of 3.47\,$\mu$m. This appears to be a genuine difference in the feature profiles between Cyg~OB2-12 and the Galactic Center. Second, it slightly underestimates the extinction in the red wing of the feature, $\lambda \gtrsim 3.55\,\mu$m. However, this is also true of the Quintuplet Cluster spectrum, and thus appears to be a shortcoming of the Gaussian fit.

Following \citet{Chiar+etal_2013}, we fit the 3.4\,$\mu$m feature with the sum of five Gaussian components

\begin{equation}
\label{eq:ch_profile}
    \Delta\tau_\lambda = \sum_{j=1}^5 \Delta\tau_j {\rm
    exp}\left[-\left(4\ln2\right) \left(\frac{\lambda-\lambda_{0,j}}
      {\Delta\lambda_j}\right)^2\right]
      ~~~,
\end{equation}
where for each component $j$, $\lambda_{0,j}$ is the central wavelength, $\Delta\lambda_j$ is the FWHM, and $\Delta\tau_j$ is the optical depth at $\lambda_0$. Note that $\Delta\lambda$ is related to $\Delta\lambda^{-1}$ via

\begin{equation}
  \Delta\lambda^{-1} = \frac{1}{\lambda_0 - \Delta\lambda/2} -
  \frac{1}{\lambda_0 + \Delta\lambda/2}
  ~~~.
\end{equation}

The best fit parameters of our Gaussian decomposition are listed in Table~\ref{table:iso_34}, and the resulting profile is presented in Figure~\ref{fig:iso34}. We have followed \citet{Chiar+etal_2013} in including components at 3.289, 3.376, 3.420, and 3.474\,$\mu$m, which they attribute to aromatic CH, the CH$_3$ asymmetric mode, the CH$_2$ asymmetric mode, and the CH$_3$ symmetric mode, respectively. While they include a 3.520\,$\mu$m component attributed to the CH$_2$ symmetric mode, we shift this component to 3.528\,$\mu$m to better fit the red wing of the feature.

The 3.3\,$\mu$m aromatic feature is best fit with $\Delta\tau_{3.3} = 0.014\pm0.005$, where the uncertainty is estimated from different treatments of the continuum. With the determination presented in Figure~\ref{fig:iso34}, the feature has roughly the same strength relative to the 3.4\,$\mu$m feature as towards the Galactic Center. 

If the 3.3\,$\mu$m feature is arising from PAH absorption, we can estimate the PAH abundance required to reproduce the observed strength. Using the absorption cross sections proposed by \citet{Draine+Li_2007}, the absorption due to PAHs $\Delta\tau_\lambda^{\rm PAH}$ is given by

\begin{equation}
    \Delta\tau_\lambda^{\rm PAH} = N_{\rm H} A_{\rm C}^{\rm PAH} \sum_j \left(\frac{2}{\pi}\right) \frac{\gamma_j \lambda_j \sigma_j}{\left(\lambda/\lambda_j - \lambda_j/\lambda\right)^2 + \gamma_j^2}
    ~~~,
\end{equation}
where for each component $j$, $\lambda_j$ is the peak wavelength, $\gamma_j\lambda_j$ is the FWHM, $\sigma_j$ is the integrated strength of the feature, and $A_{\rm C}^{\rm PAH}$ is the number of C atoms per H in PAHs. From their Table~1, the feature peaking at 3.300\,$\mu$m has $\gamma_j = 0.012$ and $\sigma_j = 3.95\times10^{-18}$\,cm per CH in neutral PAHs and $0.89\times10^{-18}$\,cm per CH in ionized PAHs.

For a column density of $N_{\rm H} = 2\times10^{22}\,$cm$^{-2}$ \citep{Oskinova+etal_2017}, the observed integrated area is compatible with $\sim 18$\,ppm of CH in neutral PAHs, or 81\,ppm of CH in ionized PAHs. The \citet{Draine+Li_2007} model (with 60\,ppm C in PAHs) has only 8\,ppm CH in neutral PAHs and $\sim 8$\,ppm CH in ionized PAHs, thus accounting for less than 50\% of the observed integrated absorption in the 3.3\,$\mu$m feature.

As noted by \citet{Chiar+etal_2013}, the Quintuplet Cluster 3.3\,$\mu$m profile is significantly wider ($\Delta\lambda \simeq 0.09\,\mu$m, $\Delta\lambda^{-1} \simeq 80$\,cm$^{-1}$) than observed in emission \citep[$\Delta\lambda \simeq 0.04\,\mu$m, $\Delta\lambda^{-1} \simeq 30$\,cm$^{-1}$;][]{Tokunaga+etal_1991,Joblin+etal_1996, Li+Draine_2001b}. The 3.3\,$\mu$m feature toward Cyg~OB2-12 appears equally broad as that observed toward the Galactic Center. Only small free-flying PAHs with $\lesssim 200$ C atoms that have been excited by single photon heating become hot enough to radiate at 3.3\,$\mu$m \citep[see][Figure~7]{Draine+Li_2007}. In contrast, the 3.3\,$\mu$m absorption feature arises from {\it all} grains. It is thus conceivable that additional PAH material is present in large grains and accounts for the observed strength of the 3.3\,$\mu$m absorption feature. Likewise, the greater diversity of material seen in absorption may explain the observed breadth relative to the emission feature. Absorption spectroscopy of the 3.3\,$\mu$m feature on more sightlines would be useful to establish whether a relatively broad extinction feature is indeed typical.

The 3.47\,$\mu$m feature is thought to arise from H atoms attached to diamond-like $sp^3$ bonded C \citep{Allamandola+etal_1992}. From detection of this feature in a sample of young stellar objects, \citet{Brooke+etal_1996} found that the feature was much better correlated with the strength of the H$_2$O ice features rather than the silicate features, and thus that the feature likely arises in dense molecular gas rather than the diffuse ISM. The detection of the feature toward the Galactic Center by \citet{Chiar+etal_2013} is consistent with this hypothesis. Thus, it is surprising that the ice-free sightline toward Cyg~OB2-12 has stronger relative absorption near 3.47\,$\mu$m than the Galactic Center sightline. If indeed this absorption is due to diamond-like carbon, then this may be a generic component of dust in the diffuse ISM. However, as illustrated by the Gaussian decomposition in Figure~\ref{fig:iso34}, other features in the vicinity of 3.47\,$\mu$m could also account for the enhanced extinction.

Assuming an absorption strength of $2.37\times10^{-17}$\,cm per CH$_3$ \citep{Chiar+etal_2013} and an integrated area of 0.55\,cm$^{-1}$ (see Table~\ref{table:iso_34}), this implies 1.2\,ppm of CH$_3$ in diamond-like form. If the 3.47\,$\mu$m feature dominates the 3.47\,$\mu$m absorption, unlike in our Gaussian decomposition, then this could be a factor of a few higher.

\subsubsection{The 6.2, 6.85, and 7.7\,$\mu$m Features}

\begin{figure}
    \centering
        \includegraphics[width=\columnwidth]{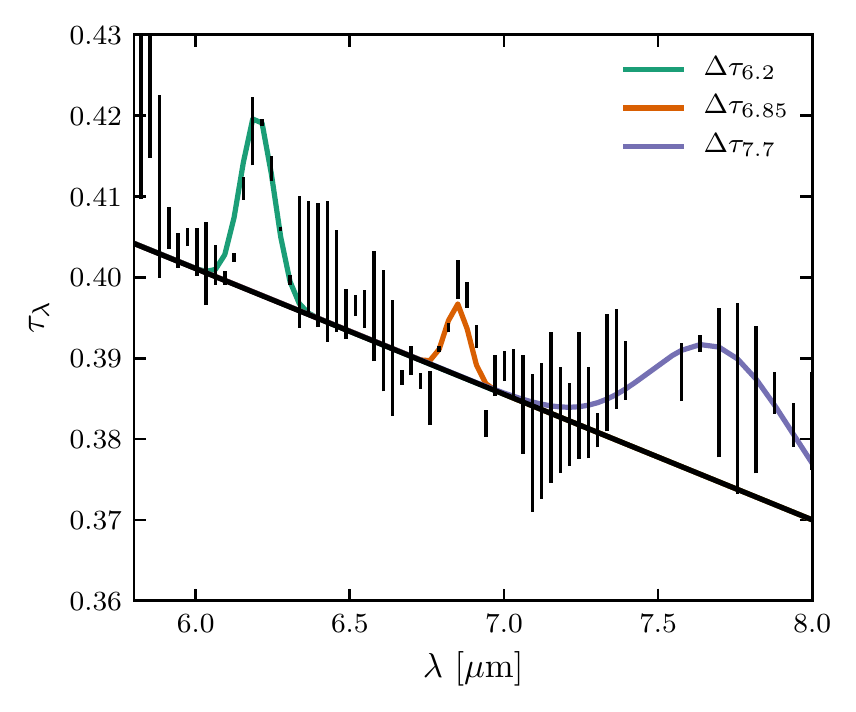}
    \caption{We perform a simultaneous fit to the 6.2, 6.85. and 7.7\,$\mu$m features observed in the {\it Spitzer} IRS data as well as the underlying continuum. The parameters of the fit Gaussian profiles are listed in Table~\ref{table:features}.} \label{fig:c_features} 
\end{figure}

\begin{figure}
    \centering
        \includegraphics[width=\columnwidth]{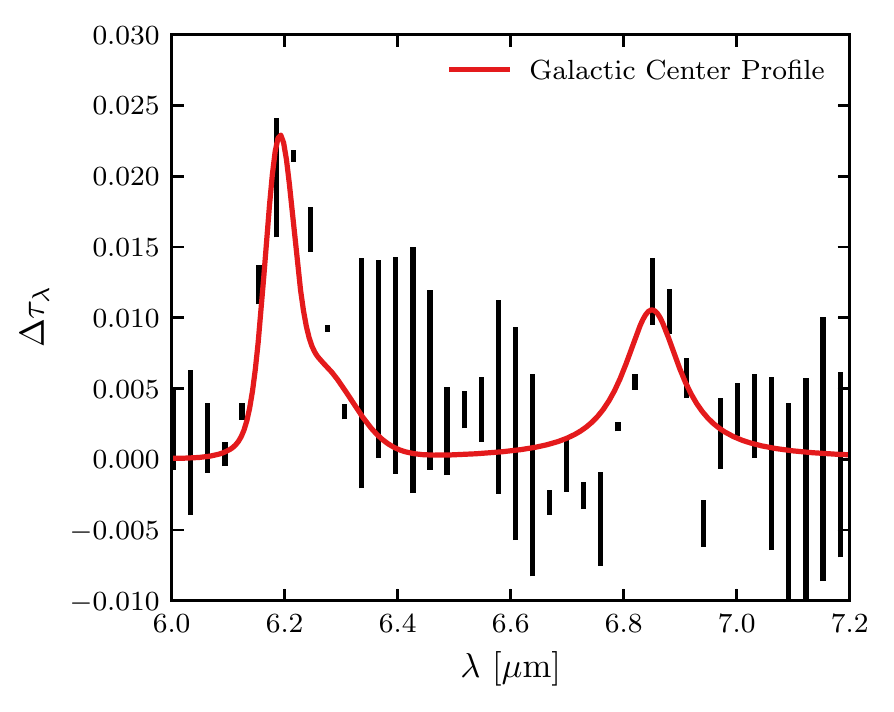}
    \caption{In red we plot the carbonaceous feature profiles derived by \citet{Chiar+etal_2000} and \citet{Chiar+etal_2013} based on observations toward the Galactic Center and rescaled to match the observed strength of the 3.4\,$\mu$m feature toward Cyg~OB2-12. We compare this to the {\it Spitzer} IRS data in black, finding that both the 6.2 and 6.85\,$\mu$m features have approximately the same  strength relative to the 3.4\,$\mu$m feature as toward the Galactic Center.} \label{fig:ch62} 
\end{figure}


The extinction curve derived from the low-resolution IRS spectrum has broad extinction features between 6 and 8\,$\mu$m (see Figure~\ref{fig:tau}). We attribute these to the 6.2 and 7.7\,$\mu$m aromatic C features and the 6.85\,$\mu$m aliphatic C feature. 

To quantify the observed strengths of the detected carbonaceous features, we adopt a Gaussian profile for each of the three features and fit the strengths and FWHMs simultaneously with the slope and intercept of a linear continuum over the wavelength range 5.8--8\,$\mu$m. No attempt is made to fit subfeatures given the limited wavelength resolution of the data. Thus, the adopted parametric model is

\begin{equation}
    \tau_\lambda = b + m\lambda + \sum_{j=1}^3 \Delta\tau_j^ {\rm
    exp}\left[-\left(4\ln2\right) \left(\frac{\lambda-\lambda_{0,j}}
      {\Delta\lambda_j}\right)^2\right]
\end{equation}
where $m$ and $b$ are the slope and intercept of the linear continuum (Equation~\ref{eq:continuum}), respectively, and for each component $j$, $\lambda_{0,j}$ is the central wavelength, $\Delta\tau_j$ is the optical depth at $\lambda_{0,j}$, and $\Delta\lambda_j$ is FWHM. We note that in this formulation, the data model is required to account for all of the 8\,$\mu$m extinction whereas the 9.7\,$\mu$m silicate feature must also be contributing at least somewhat at this wavelength. This may result in the depth of the 7.7\,$\mu$m feature and/or the continuum to be slightly overestimated.

We perform this fit using the \texttt{emcee}\footnote{\url{https://emcee.readthedocs.io/en/v2.2.1/}} Markov Chain Monte Carlo software \citep{Foreman-Mackey+etal_2013}. We adopt flat, uninformative priors on all parameters. Because of the imperfect subtraction of the Pf\,$\alpha$ line (see Section~\ref{subsec:lines}), we exclude the data from 7.4 to 7.55\,$\mu$m.

The results of the fit are presented in Table~\ref{table:features}, where the quoted uncertainties have been marginalized over all fit parameters but do not include any uncertainties inherent in the overall flux model employed to derive $\tau_\lambda$ (see Section~\ref{sec:model}), which are difficult to quantify. The best fit profiles of each of the three features are illustrated in Figure~\ref{fig:c_features}.

This model provides an excellent fit to the data over this wavelength range despite its simplicity. There is some suggestion that the Gaussian profile is overpredicting the extinction on the short wavelength side of the 6.2\,$\mu$m feature, and the continuum appears high relative to a few points near 6.7\,$\mu$m, but there is no clear evidence of unmodeled features at the sensitivity of the data.

As with the 3.4\,$\mu$m feature, we can compare the feature profiles observed toward Cyg~OB2-12 to those toward the Quintuplet Cluster \citep{Chiar+etal_2013}. In that study, the 6.2\,$\mu$m feature was divided into two distinct components. The broader of these two components was attributed to the aromatic C--C mode ($\lambda_0 = 6.25\,\mu$m, $\Delta\lambda^{-1} = 40\,$cm$^{-1}$) while the narrower component was attributed to the aliphatic C--C mode ($\lambda_0 = 6.19\,\mu$m, $\Delta\lambda^{-1} = 15\,$cm$^{-1}$). The feature optical depths relative to the 3.4\,$\mu$m feature were found to be 0.40 and 0.15, respectively. In Figure~\ref{fig:ch62}, we scale the Quintuplet Cluster profile to the $\Delta\tau_{3.4} = 0.044$ observed toward Cyg~OB2-12. The agreement is excellent, suggesting that the 3.4 and 6.2\,$\mu$m features have comparable strengths in both dense and diffuse gas.

The 6.85\,$\mu$m feature has been observed on the sightline toward the Galactic Center with both the Kuiper Airborne Observatory \citep{Tielens+etal_1996} and ISO \citep{Chiar+etal_2000} and is attributed to CH deformation modes in aliphatic carbon. The 6.85\,$\mu$m feature toward Sgr A* is well-fit by a Lorentzian with $\Delta\tau_{6.85} = 0.05\pm0.01$ and $\Delta\lambda^{-1} = 26\,$cm$^{-1}$ \citep{Chiar+etal_2000}. The sightline toward Sgr A* has $\tau_{3.4} = 0.21\pm0.01$ \citep{Chiar+etal_2000}, and so $\Delta\tau_{6.85}/\Delta\tau_{3.4} = 0.24\pm0.05$. In Figure~\ref{fig:ch62}, we scale this profile to the $\Delta\tau_{3.4} = 0.044$ observed toward Cyg~OB2-12. As with the 6.2\,$\mu$m feature, the predicted strength matches the Cyg~OB2-12 observations. Thus, the 6.85\,$\mu$m feature appears to be a generic component of dust extinction even in the diffuse ISM. 

Unlike the 3.3\,$\mu$m feature which is dominated by neutral PAHs, the 6.2 and 7.7\,$\mu$m features arise mostly from ionized grains. Using the adopted band strengths from \citet{Draine+Li_2007} and a column density of $2\times10^{22}$\,cm$^{-2}$ \citep{Oskinova+etal_2017}, we find that a 6.2\,$\mu$m optical depth of 0.022 can be produced by 35\,ppm of C in ionized PAHs. Likewise, a 7.7\,$\mu$m optical depth of 0.017 can be produced by 28\,ppm of C in ionized PAHs. This is well within the $\sim60\,$ppm of C thought to be in PAHs. The slight difference in the predicted abundances is within the systematic uncertainties of the model fit, particularly the determination of the 7.7\,$\mu$m feature strength. The observed extinction is therefore consistent with arising from PAH absorption and may constrain the amount of PAH material present in larger grains.

To our knowledge, this is the first identification of the 7.7\,$\mu$m aromatic feature in absorption. The {\it ISO}-SWS spectrum of Sgr~A* has a clear feature in the vicinity of 7.7\,$\mu$m which \citet{Chiar+etal_2000} identify with methane ice. Given that the observed depth of the feature is only slightly less than the 6.2\,$\mu$m feature on the same sightline ($0.02$ vs. $0.05\pm0.01$), it is possible that PAH absorption rather than CH$_4$ ice is responsible. Given the absence of other ice features, it is unlikely that the 7.7\,$\mu$m feature observed toward Cyg~OB2-12 arises from solid methane, and we therefore identify it as PAH absorption.

In principle, PAH absorption features at longer wavelengths are present in the {\it Spitzer} IRS spectrum. However, owing to the relative weakness of these features and the prominence of the silicate features, we find no evidence of other PAH absorption features. For instance, the \citet{Draine+Li_2007} PAH absorption profile predicts $\Delta\tau \lesssim 0.009$ in the vicinity of 8.6\,$\mu$m, which would not be discernible in these data particularly given the contribution from the 9.7\,$\mu$m feature at this wavelength.

\subsubsection{The 7.25\,$\mu$m Aliphatic Hydrocarbon Feature}

An absorption feature at 7.25\,$\mu$m associated with CH$_3$ symmetric deformation modes has been found toward Sgr~A* \citep{Chiar+etal_2000}, Seyfert~2 nuclei \citep{Dartois+etal_2004}, and luminous infrared galaxies \citep{Dartois+MunozCaro_2007} at roughly half the strength of the 6.85\,$\mu$m feature. As illustrated in Figure~\ref{fig:c_features}, the IRS data have no suggestion of a feature at this wavelength. To test this in detail, we redo our simultaneous fit of the 6.2, 6.85, and 7.7\,$\mu$m features and linear continuum with the addition of a fourth feature at 7.25\,$\mu$m having fixed $\Delta\lambda = 0.10$\,$\mu$m, consistent with the Sgr~A* sightline \citep{Chiar+etal_2000}. 

We find $\Delta\tau_{7.25} = -0.007\pm0.010$ with an upper limit of $\Delta\tau_{7.25} < 0.007$ at 95\% confidence. The inclusion of this feature has little effect on the best fit profiles of the other features, although a large $\Delta\tau_{7.25}$ would require alteration of the 7.7\,$\mu$m feature profile. Thus, while disfavored, an absorption feature at 7.25\,$\mu$m subdominant to the 6.85\,$\mu$m feature ($\Delta\tau_{6.85} = 0.009\pm0.001$) cannot be completely ruled out.

\subsubsection{The 11.53\,$\mu$m Graphite Feature}
Graphite has long been a candidate constituent of interstellar dust for its ability to produce an extinction feature consistent with the 2175\,\AA\ bump \citep{Stecher+Donn_1965}. \citet{Draine_1984} and \citet{Draine_2016} discussed an out of plane lattice resonance in polycrystalline graphite at 11.53\,$\mu$m. The expected 0.014\,$\mu$m FWHM of this feature is well-matched to the resolution of the high-resolution IRS spectrum, but no evidence of enhanced absorption is present at this wavelength, as shown in Figure~\ref{fig:rlines2}.

The most pronounced feature in this region of the spectrum is the 11.537\,$\mu$m 15--9 hydrogen recombination line. In order to constrain the strength of a possible graphite feature, over the wavelength range 11.43--11.63\,$\mu$m we model simultaneously the contribution of the 15--9 line to the total flux, a linear continuum contribution to $\tau_\lambda$, and the graphite feature at fixed $\lambda_0 = 11.53\,\mu$m and FHWM 0.014\,$\mu$m. We find that the graphite optical depth $\Delta\tau_{11.53} < 0.03$ at 95\% confidence. Assuming an opacity $\Delta\kappa = 470\,$cm$^2$\,g$^{-1}$ \citep{Draine_2016}, this implies $< 160$\,ppm of C in graphite. Unfortunately, the weakness of the feature and the presence of the recombination line prevent more stringent constraints.

\section{Discussion}
\label{sec:discussion}

The heavily reddened sightline toward Cyg~OB-12 is ideal for studying MIR extinction from dust in the diffuse ISM, and the {\it ISO}-SWS and {\it Spitzer} IRS spectra provide high-sensitivity characterization of both the spectroscopic extinction features in this wavelength range as well as the underlying continuum. Thus, the MIR extinction curve constructed in this work provides a new benchmark for models of interstellar dust.

The widely-used ``astrosilicate'' proposed by \citet{Draine+Lee_1984} included a 9.7\,$\mu$m silicate feature based on observations of dust emission in the Trapezium region \citep{Forrest+Gillett+Stein_1975}. However, the Trapezium profile FWHM $\simeq 3.45\,\mu$m is significantly broader than Cyg~OB2-12 profile derived in this work (FWHM = $2.23\pm0.01\,\mu$m) and elsewhere \citep{Roche+Aitken_1984,Bowey+Adamson+Whittet_1998}, as well as other sightlines that probe the diffuse ISM \citep{vanBreemen+etal_2011}. Thus, for use on diffuse sightlines, the astrosilicate dielectric function should be revised to accord with the Cyg~OB2-12 profile.

While the silicate and PAH features were accounted for in the astrosilicate + graphite + PAH modeling paradigm of \citet{Draine+Li_2007}, the features from aliphatic hydrocarbons at 3.4 and 6.85\,$\mu$m were not. Observations of the Cyg~OB2-12 sightline provide a detailed characterization of the feature profiles as well as their strengths relative to both the continuum extinction and the other spectroscopic features. These too should be incorporated in models of dust in the diffuse ISM.

Earlier studies of MIR extinction \citep[e.g.,][]{Bertoldi+etal_1999,Rosenthal+Bertoldi+Drapatz_2000,Hennebelle+etal_2001} suggested a pronounced minimum near 7\,$\mu$m, but the extinction curve found in this work is relatively flat in the 4--8\,$\mu$m wavelength range, consistent with recent determinations on other sightlines \citep[e.g.,][]{Lutz+etal_1996,Lutz_1999,Jiang+etal_2003,Indebetouw+etal_2005,Chapman+etal_2009,Wang+etal_2013,Xue+etal_2016}. In light of this emerging consensus on the behavior of the Galactic extinction curve at these wavelengths, dust models require significant revision. \citet{Wang+Li+Jiang_2015} suggest, for instance, that the additional absorption required could be produced by $\mu$m-sized graphite grains.

We note that the ability to measure PAH absorption on this sightline is particularly valuable since, unlike emission, absorption does not depend on the details of grain heating. Thus, the PAH optical properties are more directly accessed. High resolution followup of these features and deep searches for the longer wavelength features can test models in detail, including ionization fractions and the relative strengths of the various vibrational modes. Such a search should be possible with the Mid-Infrared Instrument on the {\it James Webb Space Telescope}.

The most uncertain aspect of the MIR emission from Cyg~OB2-12 is the contribution from the stellar wind. In particular, it remains unclear how much of the observed radio emission is the result of the collision of the Cyg~OB2-12 wind with that of its nearby companion. Very high angular resolution characterization of the wind morphology, finer than the 60\,mas separation between the stars \citep{Caballero-Nieves+etal_2014}, would be immensely valuable in understanding the origin of the radio emission and its connection to the MIR emission.

\section{Conclusions}
\label{sec:conclusions}

The principal conclusions of this work are as follows:

\begin{itemize}
    \item We develop a model for the MIR emission from Cyg~OB2-12 and its stellar wind to derive the {\it total} extinction on this sightline from {\it ISO}-SWS and {\it Spitzer} IRS spectroscopy.
    \item We identify and characterize more than sixty spectral lines, many of which are H recombination lines seen in both emission and absorption, which may help constrain models of the stellar wind.
    \item We determine the 3.4\,$\mu$m feature profile on this sightline, finding overall close agreement with the feature profile toward the Galactic Center. The extinction in the vicinity of 3.47\,$\mu$m is enhanced relative to the Galactic Center sightline, which may point to the presence of diamond-like carbon in the diffuse ISM.
    \item We find evidence for the 3.3\,$\mu$m aromatic hydrocarbon feature in extinction. The feature has a significantly broader profile than is typically seen in emission, in agreement with observations of this feature toward the Galactic Center \citep{Chiar+etal_2013}.
    \item We robustly detect extinction features at 6.2, 6.85, and 7.7\,$\mu$m associated with carbonaceous grain materials with relative strengths similar to those on the sightline toward the Galactic Center. The 6.2 and 7.7\,$\mu$m feature strengths are compatible with expectations from PAH absorption. To our knowledge, this is the first identification and characterization of the 7.7\,$\mu$m aromatic feature in absorption.
    \item Synthesizing our derived Cyg~OB2-12 extinction curve with the mean interstellar extinction curve of \citet{Schlafly+etal_2016}, we present a representative extinction curve of the diffuse ISM extending through the MIR. We demonstrate that extension of this curve into the FIR is fully compatible with dust opacities inferred from measurements of FIR emission.
\end{itemize}

\acknowledgments
{It is a pleasure to thank Paco Najarro and Eddie Schlafly for many helpful discussions on Cyg~OB2-12 and interstellar extinction. We also thank Simon Clark, Shane Fogerty, Charles Poteet, Karin Sandstrom, J.D. Smith, and Doug Whittet for stimulating conversations. This work was supported in part by NSF grants AST-1408723 and AST-1908123. This work is based on observations made with the Spitzer Space Telescope, which is operated by the Jet Propulsion Laboratory, California Institute of Technology under a contract with NASA.}

\facility{ISO, Spitzer}

\software{Astropy \citep{Astropy,Astropy_2}, \texttt{emcee} \citep{Foreman-Mackey+etal_2013}, Matplotlib \citep{Matplotlib}, NumPy \citep{NumPy}, SciPy \citep{SciPy}}

\bibliography{mybib}

\end{document}